\documentclass{aa}  
\usepackage{graphicx}
\usepackage{color}
\usepackage{url}
\usepackage{hyperref}
\usepackage[normalem]{ulem}
\usepackage{txfonts}
   \titlerunning{High-energy characteristics of \psrtar}
\usepackage{natbib}
\newcommand{\psrtar}{IGR\,J17591$-$2342}
\newcommand{\igrnew}{IGR\,J17503$-$2636}
\newcommand{\igrweak}{IGR\,J17062$-$6143}
\newcommand{\hetesrc}{HETE\,J1900.1$-$2455}
\newcommand{\maxisrc}{MAXI\,J0911$-$655}
\newcommand{\xtesrc}{XTE\,J1807$-$294}
\newcommand{\igrswing}{IGR\,J18245$-$2452}
\newcommand{\igrone}{IGR\,J00291$+$5934}
\newcommand{\xss}{XSS\,J12270$-$4859}
\newcommand{\firstpsr}{FIRST\,J102347.6$+$003841}
\newcommand{\psrjswone}{PSR\,J1023$+$0038}
\newcommand{\psrjswtwo}{PSR\,J1824$-$2452I}
\newcommand{\psrjswthree}{PSR\,J1227$-$4853}

\newcommand{\gtap}{\mathrel{\hbox{\rlap{\lower.55ex \hbox {$\sim$}}
                   \kern-.3em \raise.4ex \hbox{$>$}}}}
\newcommand{\ltap}{\mathrel{\hbox{\rlap{\lower.55ex \hbox {$\sim$}}
                   \kern-.3em \raise.4ex \hbox{$<$}}}}

\newcommand{\nustar}{{\it NuSTAR\/}}
\newcommand{\nicer}{{NICER\/}} 
\newcommand{\cxo}{{\it Chandra\/}}
\newcommand{\xmm}{{\it XMM-Newton\/}}
\newcommand{\swift}{{\it Swift\/}}
\newcommand{\Integ}{{INTEGRAL\/}} 

\def\compps{{\sc compPS}} 
\def\nthcomp{{\sc Nthcomp}}

\def\be{\begin{equation}} 
\def\ee{\end{equation}}

\begin{document}
   \title{
   High-energy characteristics of the accretion-powered millisecond pulsar \psrtar\ during its 2018 outburst}
   \titlerunning{High-energy characteristics of \psrtar}
   \authorrunning{Kuiper et al.}

   \subtitle{An XMM-Newton, NICER, NuSTAR and INTEGRAL view\\ of the 0.3--300 keV X-ray band}

   \author{L. Kuiper\inst{1}
           \and
           S. S. Tsygankov\inst{2,3}
           \and
           M. Falanga\inst{5,6}
           \and
           I. A. Mereminskiy\inst{3}
           \and
           D. K. Galloway\inst{7,8}
           \and
           J. Poutanen\inst{2,3,4}
           \and
           Z. Li\inst{9}
          }
   \offprints{L. Kuiper}

   \institute{SRON-Netherlands Institute for Space Research, Sorbonnelaan 2, 
              3584 CA, Utrecht, The Netherlands\\
              \email{L.M.Kuiper@sron.nl}
              \and
              Tuorla Observatory, Department of Physics and Astronomy, University of Turku, FI-20014, Finland
              \and
              Space Research Institute of the Russian Academy of Sciences, Profsoyuznaya Str. 84/32, 117997 Moscow, Russia
              \and
              Nordita, KTH Royal Institute of Technology and Stockholm University, Roslagstullsbacken 23, SE-10691 Stockholm, Sweden
              \and
              International Space Science Institute (ISSI), Hallerstrasse 6, 3012 Bern, Switzerland
              \and
              International Space Science Institute Beijing, No.1 Nanertiao, Zhongguancun, Haidian District, 100190 Beijing, PR China
              \and
              School of Physics and Astronomy, Monash University, Australia, VIC 3800, Australia
              \and
              Monash Centre for Astrophysics, Monash University, Australia, VIC 3800, Australia
              \and
              Department of Physics, Xiangtan University, Xiangtan, 411105, P.R. China
              }

   \date{Received xx Feb 2020 / Accepted xxx 2020}

  \abstract{
  \psrtar\ is a recently \Integ\ discovered accreting millisecond X-ray pulsar that went into outburst around July 21, 2018. To better understand the physics acting in these systems during the outburst episode we performed detailed temporal-, timing- and spectral analyses across the 0.3--300 keV band using data from \nicer, \xmm, \nustar\ and \Integ. The hard X-ray 20--60 keV outburst profile covering $\sim 85$ days is composed of four flares. During the maximum of the last flare we discovered a type-I thermonuclear burst in \Integ\ JEM-X data, posing constraints on the source distance. We derived a distance of $7.6\pm0.7$ kpc, adopting Eddington luminosity limited photospheric radius expansion burst emission and assuming anisotropic emission. In the timing analysis using all \nicer\ 1--10 keV monitoring data we observed a rather complex behaviour starting with a spin-up period (MJD 58345--58364), followed by a frequency drop (MJD 58364--58370), a episode of constant frequency (MJD 58370-58383) and concluding with irregular behaviour till the end of the outburst. The 1--50 keV phase distributions of the pulsed emission, detected up to $\sim$ 120 keV using \Integ\ ISGRI data, was decomposed in three Fourier harmonics showing that the pulsed fraction of the fundamental increases from $\sim 10\%$ to $\sim 17\%$ going from $\sim 1.5$ to $\sim 4$ keV, while the harder photons arrive earlier than the soft photons for energies $\ltap 10$ keV. The total emission spectrum of \psrtar\ across the 0.3--150 keV band could adequately be fitted in terms of an absorbed \compps\ model yielding as best fit parameters a column density of $N_{\rm H}=(2.09\pm0.05)\times10^{22}$cm$^{-2}$, a blackbody seed photon temperature $kT_{\rm bb, seed}$ of $0.64\pm 0.02$ keV, electron temperature $kT_{\rm e}=38.8\pm 1.2$ keV and Thomson optical depth $\tau_{\rm T}=1.59\pm 0.04$.
  The fit normalisation results in an emission area radius of $11.3\pm 0.5$ km adopting a distance of 7.6 kpc. Finally, the results are discussed within the framework of accretion physics- and X-ray thermonuclear burst theory.
  }
   
   \maketitle
%

\section{Introduction}
\label{intro}
Accreting millisecond X-ray pulsars (AMXPs) form a subset of low-mass X-ray binaries (LMXBs) that exhibit coherent millisecond pulsations during outburst,  with spin periods ranging from 1.7 to 9.5 ms (based on the current sample
of nineteen members). As with some other LMXBs, the AMXPs typically exhibit transient outbursts as often as every few years, reaching luminosities
of $10^{36-37}$~erg~s$^{-1}$ in the X-ray band.
The outburst duration is typically a couple of weeks, however, there are a few exceptions, e.g., \xtesrc\ \citep{linares05}, \hetesrc\ \citep{simon18} 
and \maxisrc\ \citep{parikh19}, which showed outburst durations of $\ga 150$ days, $\sim10.4$ yr and $\sim3.2$ yr, respectively. 

It is believed that these AMXPs are the progenitors of rotation-powered millisecond pulsars (RPMPs) which are mainly detected at radio frequencies and/or at gamma-ray
energies, and manifest either as a member of a binary system or as solitary stars. The underlying recycling scenario \citep{Alpar82,r82} assumes 
that during the accretion phase mass with large specific angular momentum is transferred towards a slowly rotating neutron star, which spins-up by the 
resulting accretion torques. It is expected that at the end of the active accretion phase a millisecond pulsar turns on shining from the radio- to the gamma-ray
regime powered by the rotation of its magnetic field.

The link between RPMPs and AMXPs has recently been established by the detection of three so-called ``transitional'' pulsars \firstpsr\ (\psrjswone), \igrswing\ (\psrjswtwo) and \xss\ (\psrjswthree), which swing back from one state to the other state and vice versa \citep[see e.g.][the latter three citations are for \xss]{archibald09,papitto13c,demartino10,bassa14,Roy15}, respectively.

Currently, nineteen members belonging to the AMXP class are discovered that went into full outburst (including our target \psrtar\ and transitional ms-pulsar 
\igrswing). This sample excludes, however, the other two transitional pulsars \psrjswone\ and \psrjswthree, which both have not (yet) undergone a full outburst, 
and LMXB Aql X-1, which showed (intermittent) pulsed emission only during a small $\sim 120$ s interval. More information on the characteristics of this intriguing 
 source class can be found in reviews by \citet{Wijnands2006}, \citet{patruno12} and \citet {Campana18}.

To understand the evolution from the AMXP-stage to the RPMP-stage better it is important to perform detail studies of the accretion-related phenomena in these transient 
systems, which spend most of the time in quiescence, through temporal, (pulse) timing and spectral analyses of X-ray data collected across an as wide as possible energy band.

In the last decade(s) {INTErnational Gamma-Ray Astrophysical Laboratory}  \citep[\Integ,][]{winkler03} turned out to be an efficient observatory to discover new members, eight in total now, of the AMXP class.
This work focusses on a new member of the AMXP class, \psrtar, discovered by \Integ\ on 2018 Aug 12 \citep{ducci2018}. 

Soon after its discovery the soft X-ray counterpart was detected by the XRT on-board of the \textit{Neil Gehrels Swift Observatory} \citep{Bozzo18} yielding an improved location, followed by
the identification as a 1.9 ms AMXP with a 8.8 h orbital period using \nicer\ and \nustar\ observations \citep{ferrigno2018}. Also, at radio frequencies 
(ATCA; 5.5 and 9 GHz) and in the near-infrared (NIR) band (K$_{\rm s}$ and H-bands; HAWK-I UT4 of the VLT) \psrtar\ was detected \citep[see][respectively]{Russell2018,shaw2018}. 

Interestingly, examination of archival \swift\ BAT data (15--150 keV) showed that \psrtar\ went off about 20 days before the discovery by \Integ\ \citep{krimm2018}, which 
could not observe the Galactic center region earlier because of observational constraints. Finally, the early findings obtained during activated ToO programs with the 
High-Energy transmission Grating Spectrometer aboard \cxo\ \citep{NowakATEL18} and with \Integ\ and \nustar\ \citep{KuiperATEL18}, were reported on, respectively.

From X-ray timing measurements \citep{Sanna2018} and optical/NIR VLT X-shooter observations \citep[see e.g. Sect. 2.4 of][]{Nowak2019} the nature of the companion star 
could be determined. It could be classified consistently as a late-spectral type star with age between 8 and 12 Gyr and a mass in the range 0.85--0.92 $M_{\sun}$, assuming a 1.4 $M_{\sun}$ mass for the X-ray pulsar. In this case the inclination of the orbit is rather low and in the range 28\degr--30\degr.

In this paper we present the broad-band (0.3--300 keV) results using all available \Integ-ISGRI (20--300 keV) data, including two ToO observations, \nustar\ (3--79 keV) ToO observations and one \xmm\ (0.3--15 keV) ToO observation supplemented by \nicer\ (0.2--12 keV) monitoring observations in order to shed light on the physical processes acting in this AMXP during its 2018 outburst, and on the distance and the hydrogen column density $N_{\rm H}$ to the system. 

\section{Instruments and observations}

\begin{figure*}[t]
  \centering
  \includegraphics[width=16.0cm]{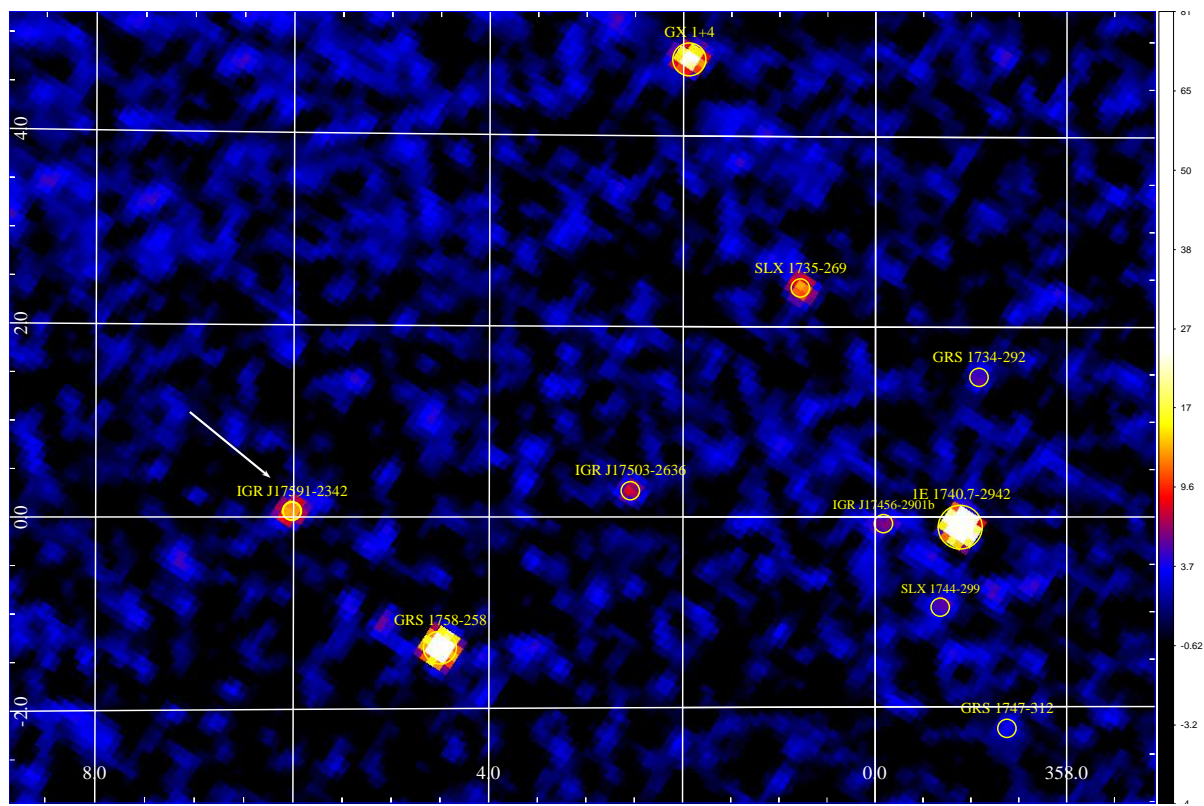}
  \caption{INTEGRAL ISGRI 20--60 keV significance map in Galactic coordinates for Rev-1986 (MJD 58340.661-58341.806) of the sky region showing the new 
           hard X-ray transient \psrtar, indicated by an arrow, at a significance of $\sim 13.8\sigma$. Other detected hard X-ray 
           sources during this period are indicated as well among them, \igrnew, which was detected for the first time in JEM-X data during the 
           Galactic bulge observations of Rev-1986 \citep{chenevez2018}.
 }
  \label{isgri_map}
\end{figure*}


\subsection{INTEGRAL}
\label{instr_integral}
The INTEGRAL spacecraft \citep{winkler03}, launched on 2002 October 17, has two main 
$\gamma$-ray instruments aboard: a high-angular-resolution imager IBIS \citep{ubertini03} and
a high-energy-resolution spectrometer SPI \citep{vedrenne03}. To provide complementary observations 
in the X-ray and optical energy bands the payload is further equipped by two monitor instruments:
the Joint European Monitor for X-rays \citep[JEM-X; ][]{lund2003} and the Optical Monitoring Camera 
(OMC; 500--600 nm). To enable image reconstruction in the hard X-ray/soft $\gamma$-ray band the 
high-energy instruments make use of coded aperture masks.
 
The INTEGRAL Soft-Gamma Ray Imager \citep[ISGRI; the upper detector system of IBIS,][]{lebrun03} with its $19\degr \times 19\degr$ field of 
view (50\% partially coded) has an angular resolution of about $12\arcmin$ and could locate and separate 
high-energy sources at an accuracy of better than $1\arcmin$ (for a $>10\sigma$ source) in crowded fields with an 
unprecedented sensitivity ($\sim$ 960 cm$^2$ at 50 keV).
Its energy resolution of about 7\% at 100 keV is amply sufficient to determine the (continuum) spectral 
properties of hard X-ray sources in the $\sim$ 20--500 keV energy band.
The timing accuracy of the ISGRI time stamps recorded on board is about $61\mu$s. 

JEM-X consists of two identical telescopes each having a field of view of $7\fdg5$ (diameter) at half response 
and is able to pinpoint a $15\sigma$ source with a 90\% location accuracy of about $1\arcmin$. 
Its timing accuracy ($3\sigma$) and energy resolution are $122.1\,\mu$s and 1.3 keV at 10 keV, respectively. 
In this work, guided by sensitivity considerations, we used only data recorded by ISGRI, sensitive 
to photons with energies in the range $\sim$20 keV -- 1 MeV, and JEM-X operating in the 3--35 keV 
X-ray band. 

In its default operation mode INTEGRAL observes the sky in a dither pattern with $2\degr$ steps to suppress systematical effects, which could be 
either rectangular (e.g. a $5 \times 5$ dither pattern with 25 grid points) or hexagonal (with 7 grid points; target in the middle). 
Typical integration times for each grid point (pointing/sub-observation) are in the range 1800--3600 seconds. This strategy drives the structure 
of the INTEGRAL data archive which is organised in so-called science windows (Scw; $\equiv$ one grid point) per INTEGRAL orbital revolution (Rev), which currently lasts for 
about 2.67 days, containing the data from all instruments for a given pointing. 
Most of the INTEGRAL data reduction in this study was performed with the Offline 
Scientific Analysis (OSA) version 10.2 and 11.0 distributed by the INTEGRAL Science Data Centre 
\citep[ISDC; see e.g.][]{courvoisier03}.

Shortly after the announcement of a new high-energy transient, \psrtar, in the \Integ-ISGRI Galactic Center observations 
during satellite revolution 1986 (2018 Aug 10--11)  \citep{ducci2018,ferrigno2018} 
we activated our coordinated \Integ, \nustar\ and \xmm\ target-of-opportunity program 
to study \psrtar\ in detail. 
In particular, \Integ\ performed observations in hexagonal dither-mode during revolutions 1989 and 1992, lasting from August 17 14:03:53 to August 19 18:08:49 2018, 
and August 25 17:30:47 to August 27 17:34:10 of 2018, respectively. 
For our study we used all available \Integ\ observations with \psrtar\ within $14\fdg5$ from the IBIS pointing axis, from 
its detection in Rev-1986 up to and including Rev-2016.
This set includes data obtained in the course of observation programs 1520001 (Galactic bulge monitoring; PI E. Kuulkers),
1520022 (Continued Observation of the Galactic Center Region with INTEGRAL; PI J. Wilms), 1520038 (Galactic Center field:
deep exposure; PI S.A. Grebenev) and  1420021 (Broad view on high energy Galactic background: Galactic Center; PI R.A. Sunyaev).
We screened the resulting dataset further and excluded events registered during time periods where the ISGRI count-rate behaved 
erratically e.g. near perigeum ingress/egress.

We performed a standard \Integ\ ISGRI imaging (and subsequent mosaicking) analysis \citep{gold03}, using the imaging 
S/W tools embedded in OSA 10.2, of all available Galactic Center observations of Rev-1986 adopting several broad energy bands.
The significance map for the 20--60 keV band in Galactic coordinates, combining these Rev-1986 Galactic Center observations, 
is shown in Fig. \ref{isgri_map}. It covers a $12\degr \times 8\degr$ region encompassing the new transient \psrtar, which is detected 
at a $13.8\sigma$ confidence level and marked by a white arrow.
Other high-energy sources are indicated as well, including the (also in Rev-1986, JEM-X detected) new transient \igrnew\ \citep{chenevez2018}.
LMXB GX 5$-$1, located at $(l,b)=(+5.0802,-1.0234)$, i.e. within $1\fdg5$ of \psrtar, is hardly detectable at energies above 20 keV; however, 
it shines brightly in the canonical $\sim$2--10 keV X-ray band, and so causes straylight complications in the analysis of the \nustar\ ToO 
observations discussed in Sect. \ref{instr_nustar}.

Mosaics combining \Integ-ISGRI data from Rev-1986 up to and including Rev-2006 (Aug. 10 -- Oct. 2, 2019), have also been produced 
for several broad energy bands. The maps showed that soft gamma-ray emission from \psrtar\ could be detected up to $\sim$ 150 keV, with 
a $8.2 \sigma$ significance in the 120--150 keV band. Above, in the 150--300 keV band no source signal could be found at the location of \psrtar. 
Finally, we performed timing analysis  of the ISGRI data outside the OSA environment and selected only those events from non-noisy 
detector pixels having a pixel illumination factor of at least 25\% and with rise times between channels 7 and 90 \citep{lebrun03}.

\begin{figure*}[t]
  \centering
  \includegraphics[angle=90,width=16.0cm]{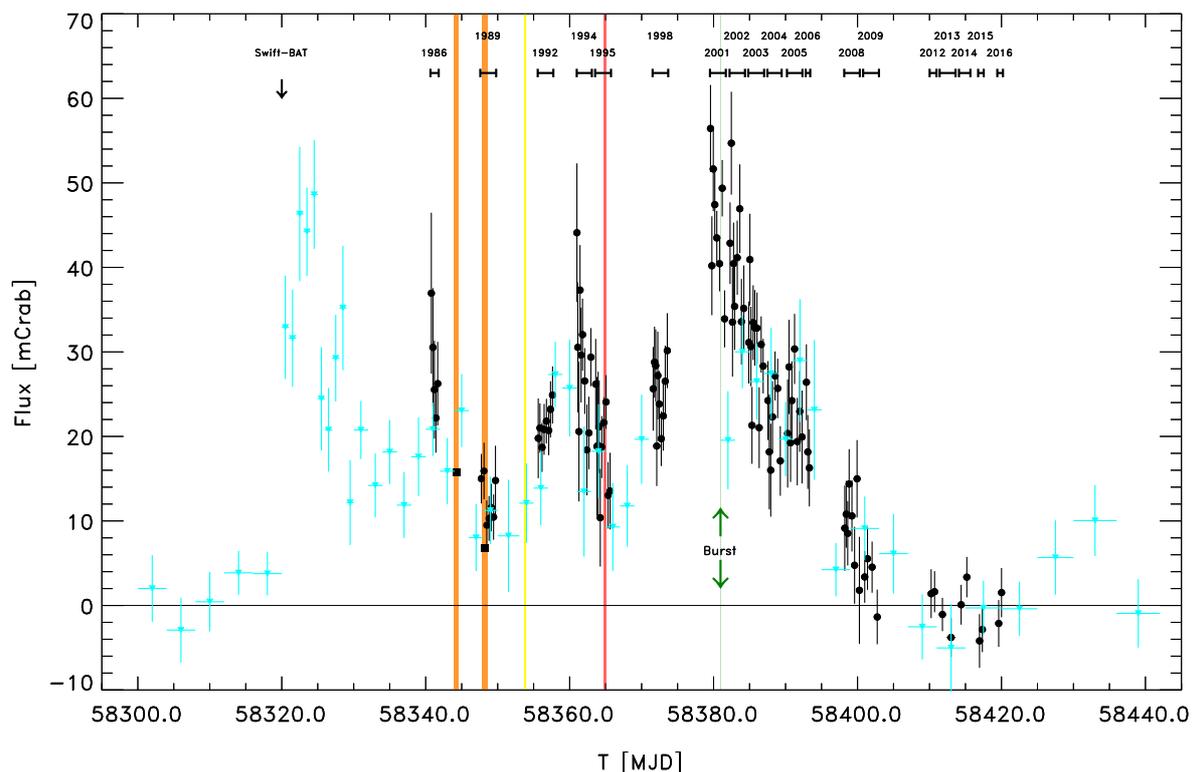}
  \caption{Hard X-ray outburst profile of \psrtar\ for the 20--60 keV band (\Integ\ ISGRI; black data points). The flux is expressed in mCrab. 
  \swift\ BAT measurements for the 15--50 keV band are superposed (cyan colored triangles; expressed in mCrab units as well). 
  The onset of the 2018-outburst of \psrtar\ starts around MJD 58320
  and ends at MJD 58405, lasting for about 85 days. Multiple flares are visible, and during the one that reached maximum luminosity 
  we detected a bright burst at the end of \Integ\ Rev-2001, indicated by green arrows/line. The orange colored bands indicate the time intervals 
  of the two \nustar\ ToO observations, while the red and yellow bands do so for the \xmm\ and \cxo\ ToO's, respectively. At the top of the panel 
  the timeline of the \Integ\ ISGRI Galactic Center observations are denoted by their corresponding revolution identifier. The onset of the outburst, 
  20 days before the actual discovery of \psrtar\ by \Integ, as identified in \swift\ BAT data \citep{krimm2018}, is indicated as well.
 }
  \label{outburst_profile}
\end{figure*}

\subsection{NICER}
\label{instr_nicer}
The Neutron star Interior Composition ExploreR (\nicer), launched on 2017 June 3, is an International Space Station payload dedicated to (spectral) timing analysis studies
in the 0.2--12 keV band at an unprecedented time resolution of $\sim 100$ ns with respect to UTC using GPS synchronization \citep{arzoumanian2014}. 
It consists of 56 X-ray concentrator (XRC) and matching silicon drift detector (SDD) pairs, forming in combination the X-ray Timing Instrument (XTI), that 
registers photons at a $\sim150$ eV spectral resolution at 6 keV from a $6\arcmin$-diameter field-of-view.
The effective area in the 1--2 keV band is about twice as large as that of the \xmm\ EPIC-pn. All these characteristics make this non-imaging
instrument an excellent tool to monitor (transient) X-ray sources at high time resolution.

\nicer\ started regular observations of \psrtar\ on 2018 Aug 14  (MJD 58345; Obs. id. 1200310101), a couple of days after the detection by \Integ\, and ceased on 
2018 Oct 17 (MJD 58408; Obs. id. 1200310139), when the source reached undetectable levels. The total on-source exposure time (37 observations combined; screened
for background flares) was 93.41 ks. During the last two observations (Obs. ids. 1200310138/1200310139) performed on 2018 Oct 16/17, respectively, for a total
combined exposure of 5.455 ks \psrtar\ was not detectable anymore, and so we used these observations to estimate the local background.
In this work the \nicer\ data were mainly used to construct accurate timing models (ephemerides) for the rotational- and orbital motion of the neutron star in \psrtar, 
and to study the pulse profile morphology and (background subtracted) pulsed fraction as a function of energy.

\subsection{NuSTAR}
\label{instr_nustar}
The Nuclear Spectroscopic Telescope Array (\nustar), launched on 2012 June 13, operates in the 3--79 keV band and provides for the first time 
fine imaging (angular resolution $18\arcsec$ FWHM) for energies above 10 keV through its two depth-graded multilayer-coated Wolter-I (conical approximation) 
X-ray optics, which focus the hard X-rays onto two independent Focal Plane Modules (FPMA and FPMB), each composed of four solid state CdZnTe pixel detector 
arrays, separated from the optics at a $\sim$10 m focal length via a deployable mast \citep{harrison2013}. Its field-of-view is about $10\arcmin$ at 10 keV 
and decreases gradually at higher energies. The energy resolution $\Delta E/E$ is about 4\% at 10 keV and 1\% at 68 keV.

The focal plane instruments record events with a $2\,\mu$s precision. These intrinsic time stamps are related to an on-broad time reference system composed of a 24 MHz 
temperature compensated crystal oscillator. Next, the space-craft clock is correlated routinely with ground stations clocks having their own standards.
The net effect of the correlation processes is that the original time tag of the recorded events is considerably blurred and is only accurate at the 2--3 ms level in UTC
time system \citep{madsen2015a}, too coarse to perform detailed timing analysis for AMXPs. The main cause of the blurring originates from the variations of the clock 
frequency due to changes in the thermal environment of \nustar\ along its 96 minutes orbit around Earth.
However, \citet{gotthelf2017} have developed a method, applicable only for sources which are `strong' enough ($\sim75$ counts per \nustar\ orbit) and with known 
ephemeris, to recover the \nustar\ relative timing with a resolution down to $\la 15 \mu$s. 

\nustar\ observed \psrtar\ on 2018 Aug 13 for 27.3 ks (Obs. id. 90401331002; public ToO-I; MJD $\sim 58344$) and on 2018 Aug 17 for 29.4 ks (Obs. id. 80301311002; 
our activated ToO-II; MJD 58348), see also the orange color bars in Fig. \ref{outburst_profile}. 
For both observations in the subsequent analysis we used the `cleaned' event files from the default \nustar\ pipeline analysis for both FPMA and FPMB.
In particular, we barycentered these event data using {\tt HEASOFT} multi-mission tool {\tt barycorr v2.1} with \nustar\ clock-correction file \#85 (valid up to 2018 Oct 2), 
Solar System ephemeris DE405 and the optical position of \psrtar\ \citep{shaw2018}. Furthermore, we applied the correction method outlined in Sect. 2.2 of \citet{gotthelf2017}
to the timing analysis of the \nustar\ \psrtar\ data. In the timing analysis we used only events which are located in a circular region of radius $60\arcsec$ around the optical location of
\psrtar. To obtain `background' corrected source parameters like pulsed fraction (as function of energy) we chose a source-free circular background region located on the same chip
with a radial aperture of $60\arcsec$ centered at $(\alpha_{2000},\delta_{2000})^{bg}=(17^{\hbox{\scriptsize h}}59^{\hbox{\scriptsize m}}07\fs89,-23\degr39\arcmin16\,\farcs0)$. 

\nustar\ imaging analysis in different energy bands showed that \psrtar\ was detectable across the full \nustar\ bandpass of 3--79 keV, however, below $\sim$20 keV straylight
from GX 5$-$1 loomed up hampering spectral analysis. 
Finally, we noticed a decaying trend in the \nustar\ count rate during the first ToO observation of \psrtar, while during the 
second one the count rate was increasing.


\subsection{XMM-Newton}
\label{instr_xmm}
On 2018 Sept 3 17:59:03 \xmm\ started a 36.7 ks ToO observation of \psrtar\ (MJD 58364.784--58365.166) on our request (Obs. id. 0795750101; \xmm\ revolution-3431; 
see the red colored bar in Fig. \ref{outburst_profile}). 
The (imaging) EPIC pn instrument \citep[][0.15--12 keV]{struder2001} aboard \xmm\ was operated in Timing Mode (TM) 
allowing timing studies at $\sim 30\,\mu$s time resolution. The other (imaging) EPIC instrument equipped with two cameras based on MOS CCDs \citep{turner2001}, MOS-1 and MOS-2, 
were put in Small window and Timing uncompressed modes, respectively.
The latter mode can handle maximum count rates of up to 35 mCrab without pile-up complications, and is suitable for our spectral study
of the total emission of \psrtar. Its time resolution of $\sim 1.75$ ms, however, is insufficient to perform timing studies at ms-time scales.
The incoming X-rays were attenuated through thin filters in front of the three EPIC instruments. 

The (non-imaging) Reflection Grating Spectrometers \citep{denherder2001} on board \xmm\ operated in default mode (HighEventRate with Single Event Selection; HER + SES), collecting 
spectral information in the $\sim 0.35$ -- $2.5$ keV band. We used only spectral data dispersed into the first order.

We run the {\tt XMM-SAS} pipeline analysis scripts (ODS-version 17.0; 2018-06-20) for all five \xmm\ (X-ray) instruments. The \xmm\ EPIC pn data were subsequently barycentered using the 
{\tt SAS barycen 1.21} script adopting Solar System ephemeris DE405 and the optical position of \psrtar\ \citep{shaw2018}. In the \xmm\ EPIC-Pn timing analysis we further select 
on the one-dimensional spatial parameter {\tt RAWX} by defining the source-region as {\tt RAWX} interval [30,44] (15 pixels of width $4\farcs1$ = 61\farcs5) and background region 
the union of {\tt RAWX} [11,19] and [55,63] (18 pixels width in combination), chosen far from the source region.

We also checked the data for the presence of soft proton background flares, but we detected none, and therefore further cleaning was not required. During the \xmm\ observation we 
noticed a (smoothly) decaying trend in the count rate of \psrtar.

\section{The outburst profile at hard X-rays}
\label{sect_outburst}
The standard OSA \Integ\ ISGRI imaging analysis yielded deconvolved sky images for each science window providing count rate, count rate variance, significance
and exposure information across the field-of-view. From these maps we extracted the count rates plus uncertainties of \psrtar\ in the 20--60 keV band, and
compared these to those of the Crab (Nebula plus pulsar) from observations, Revs. 1987, 1991, 1996, 1999, 2000 and 2010, covering our \psrtar\ observation period. 
The scatter in the Crab count rates (e.g. 20--60 keV rate: $1$ Crab $\equiv 110.4$ c\,$^{-1}$) is about 3--5\% and are negligible with respect to the \psrtar\ count rate uncertainties. 
We combined typically 5--7 data points to reach exposure times of about 10 ks. The results (black data points), based on all available \Integ\ observations of the 
Galactic Center region from Rev-1986 till 2016 (i.e. Aug 10 -- Oct 29) are shown in Fig. \ref{outburst_profile}. At the top of the plot the time line of the \Integ\ 
Galactic Center observations, labelled with their associated revolution number, is indicated.

We complemented our ISGRI measurements with data (aqua colored in Fig. \ref{outburst_profile}) from the \swift\ BAT Hard X-ray (15--50 keV) Transient Monitor program \citep{krimm2013}. The BAT data for period MJD 58300--58352 (July 1 -- August 22, 2018) have been obtained via private communication with Dr. H. Krimm. Clearly visible in Fig. \ref{outburst_profile} is that the outburst of \psrtar\ actually started near MJD 58320 \citep{krimm2018} about 20 days earlier than its detection by \Integ.

The total outburst of \psrtar\ -- with multiple flares -- lasts for about 85 days, from MJD 58320 (2018 July 21) till 58405 (2018 Oct 14), considerably longer than the typical outburst duration of 2--3 weeks for the majority of AMXPs. During the last, most prominent flare near MJD 58380, we discovered a bright type-I thermonuclear X-ray 
burst in \Integ\ JEM-X data near the end of Rev-2001 (indicated by green arrows/line in Fig.\,\ref{outburst_profile}).
Finally, the time intervals of target-of-opportunity observations by other high-energy instruments are shown: the two \nustar\ observations (orange), \xmm\ (red) and \cxo\ (yellow).

\begin{table}[t] 
{\small
\caption{Positional, orbital and rotational parameters derived in this work (or set fixed) for \psrtar.}
\centering
\begin{tabular}{lcc} 
\hline 
Parameter &Unit & Value \\
\hline 
\noalign{\smallskip}  
$\alpha_{2000}$\tablefoottext{a}          &                                 & $17^{\hbox{\scriptsize h}} 59^{\hbox{\scriptsize m}} 02\fs87$\\ 
$\delta_{2000}$\tablefoottext{a}          &                                 & $-23\degr43\arcmin08\farcs2$ \\ 
$ a_x\, \sin i$\tablefoottext{b}     &lt-s                           & $1.227\,716(8)$ \\ 
$ e $                                     &                                 & $0$\\
$ P_{orb} $\tablefoottext{c}              &s                              & $31\,684.745(1)$\\
$ T_{asc} $\tablefoottext{c}              &MJD; TDB                      & $58345.171\,978\,7(5)$\\
JPL SS-Ephemeris                          &                                 & DE405                 \\
\noalign{\smallskip}  
\hline  
\noalign{\smallskip}  
\multicolumn{3}{c}{Ephemeris-1 (Segment-1)}\\
\noalign{\smallskip}
Begin -- End                              &MJD; TDB                     & 58345 -- 58364\\
$t_0$                               &MJD; TDB                       & 58357\\
$\nu$ \tablefoottext{c}                                     &Hz            & $527.425\,700\,524(3)$\\
$\dot\nu$ \tablefoottext{c}                                 &Hz s$^{-1}$          & $(+4.07\pm 0.79)\times 10^{-14}$\\
$\Phi_0$                                  &                                 & $0.468$\\
\noalign{\smallskip}  
RMS                                       &$\mu$s                         & 26.2\\
\noalign{\medskip}  
\multicolumn{3}{c}{Ephemeris-2 (Segment-3)}\\
\noalign{\medskip}
Begin -- End                              &MJD; TDB                       & 58370 -- 58383\\
$t_0$                               &MJD; TDB                       & 58370\\
$\nu$ \tablefoottext{c}                                     &Hz            & $527.425\,700\,381(5)$\\
$\dot\nu$ \tablefoottext{c}                                &Hz s$^{-1}$         & 0\\
$\Phi_0$                                  &                                 & $0.735$\\
\noalign{\smallskip}  
RMS                                       &$\mu$s                         & 24.6\\
\noalign{\smallskip}  
\hline  
\end{tabular}
\tablefoot{ \tablefoottext{a}{Values adopted from \citet{shaw2018} and fixed in this work. Typical uncertainty is $0\farcs03$.}
            \tablefoottext{b}{Value adopted from \citet{Sanna2018} and fixed in this work.}
            \tablefoottext{c}{Errors in parameters are at $1\sigma$ confidence level as obtained in this work.}
          }
\label{table:eph} 
}
\end{table} 

\begin{figure}[t]
  \centering
  \includegraphics[angle=0,width=8.0cm,bb=70 160 550 660,clip=]{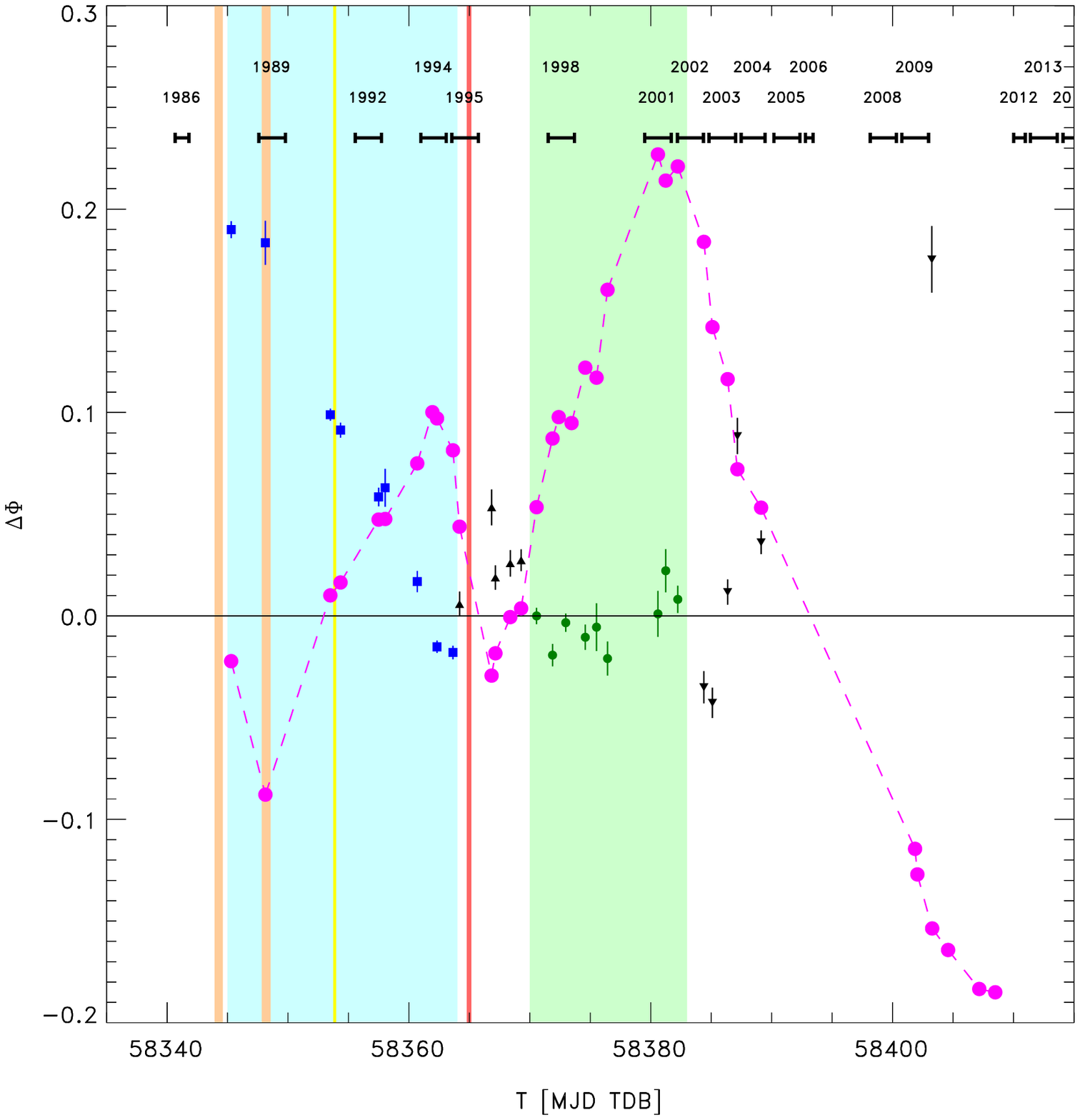}
  \caption{Timing residuals of {\em all\/} \nicer\  \psrtar\ pulse template arrival times (in phase units) with respect to the best phase-coherent
           model for segment-3, which covers MJD 58370--58383 (green band/data points). Deterministic (segment-1, covering MJD 58345--58364, blue band/data points) as well as more noisy/erratic segments are shown too. In orange, yellow and red the two \nustar, \cxo\ 
           and \xmm\ ToO's are indicated, respectively. The INTEGRAL observation timeline (with satellite revolutions labels) with \psrtar\ in the field of view are shown as well in the top part of the figure. Finally, the magenta data points and dashed line show the shape of the \nicer\ 1--10 keV daily averaged count rate. Surprisingly, phase coherent timing models can only be generated for the rising parts of the light curve.
          }
  \label{toa_residuals}
\end{figure}

\section{Timing}
\label{sect_tm}
Irrespective of the instrument, in timing analyses we have to convert the Terrestial Time (TT) arrival times of the (selected) events to 
arrival times at the solar system barycenter (in TDB time scale). Throughout in this work we have used in this process: 1) the JPL DE405 solar system ephemeris, 2) 
the instantaneous spacecraft position with respect to Earth center and 3) the (most accurate) sub-arcsecond celestial position of the NIR counterpart of \psrtar:
$(\alpha,\delta)=(17^{\hbox{\scriptsize h}}59^{\hbox{\scriptsize m}}02\fs87,-23\degr43\arcmin08\,\farcs2)$ for epoch J2000 \citep[][typical astrometric uncertainty of $0\farcs03$]{shaw2018}, which corresponds to 
$(l,b)=(+6.0213487,+0.070002)$ in Galactic coordinates (see also Fig.\,\ref{isgri_map}).

\subsection{NICER timing analysis}
\label{sect_tm_nicer}
We selected for our timing analysis `cleaned' \nicer\ XTI events from the standard pipeline analysis with measured energies between 1 and 10 keV. Events with energies between
12--15 keV, however, were used to flag periods with high-background levels (e.g. South Atlantic Anomaly ingress/egress, etc.) as bad, and these intervals were ignored
in further analysis. Moreover, events from noisy/malfunctioning detectors are ignored.
The screened events were subsequently barycentered using a (multi-instrument serving) IDL procedure.

Next, we selected the barycentered arrival times from \nicer\ observations 1200310101 -- 1200310110 (MJD 58345--58364) for the 1--10 keV band,
and applied an optimization algorithm based on a {\tt SIMPLEX} optimization scheme for four parameters simultaneously \citep[see e.g.][for the first application]{deFalcoa}, that finds the global maximum
of the $Z_3^2(\phi)$-test statistics \citep{buccheri1983} with respect to the rotational parameters $\nu, \dot\nu$ and binary orbital parameters $P_{\rm orb}$ and $T_{\rm asc}$, 
refering to the pulse frequency, frequency derivative, orbital period and time of the ascending node, respectively (the pulse phase $\phi=\phi(\nu, \dot\nu,P_{\rm orb},T_{\rm asc})$).

In this process the arrival time of each event is first corrected for the binary motion (thus involving $P_{\rm orb}$ and $T_{\rm asc}$, while keeping the eccentricity $e$ fixed at $0$ and 
projected orbital size of the neutron star orbit, $a_x\cdot \sin(i)$, at $1.227\,716$ lt-s, see \citealt{Sanna2018}) and subsequently converted into pulse phase using $\nu$ and $\dot\nu$ (the other two free parameters).
As start parameters for the optimization procedure we used the values given in column three of table 1 of \citet{Sanna2018}, i.e. the \nicer\ derived ephemeris characteristics evaluated
for the time interval of MJD 58345--58354, about two times smaller than the interval width we use.
Our derived orbital parameter values for $P_{\rm orb}$ and $T_{\rm asc}$ are consistent with those given by \citet{Sanna2018} in their table 1 (\nicer\ column), but are more accurate because of the longer baseline. 

\begin{figure}[t]
  \centering
  \includegraphics[angle=0,width=8.0cm]{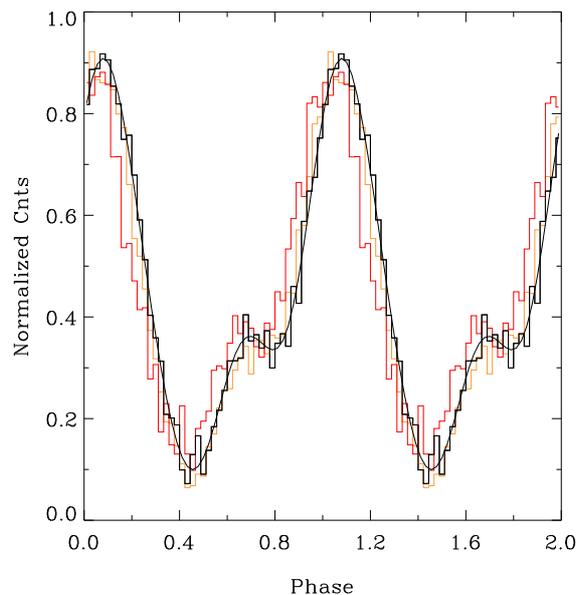}
  \caption{NICER pulse profiles of \psrtar, using data collected during segment-1 (MJD 58345--58364), for three different energy bands: 1--3 keV (including truncated Fourier series fit with 3 harmonics; black), 3--5 keV (orange) and 5--10 keV (red).
           The shift towards earlier phases of the higher energy profiles is evident.
          }
  \label{pp_nicer_vs_e}
\end{figure}

\subsection{Timing solutions: ephemerides}
\label{sect_tm_sol}
The up-to-date accurate orbital parameters (see Table \ref{table:eph}) were subsequently used in correcting the (\nicer) pulse arrival times for the orbital motion delays.
Next, we applied a so-called Time-of-Arrival (ToA) analysis yielding the most accurate estimates possible for the pulse frequency $\nu$ (and frequency first time derivative $\dot\nu$, 
if required) over a certain time stretch (validity interval). This ToA method has been described earlier in detail in Sect. 4.1 of \citet{kuiper2009}.

\begin{figure*}[t]
  \centering
  \includegraphics[angle=0,width=16.0cm]{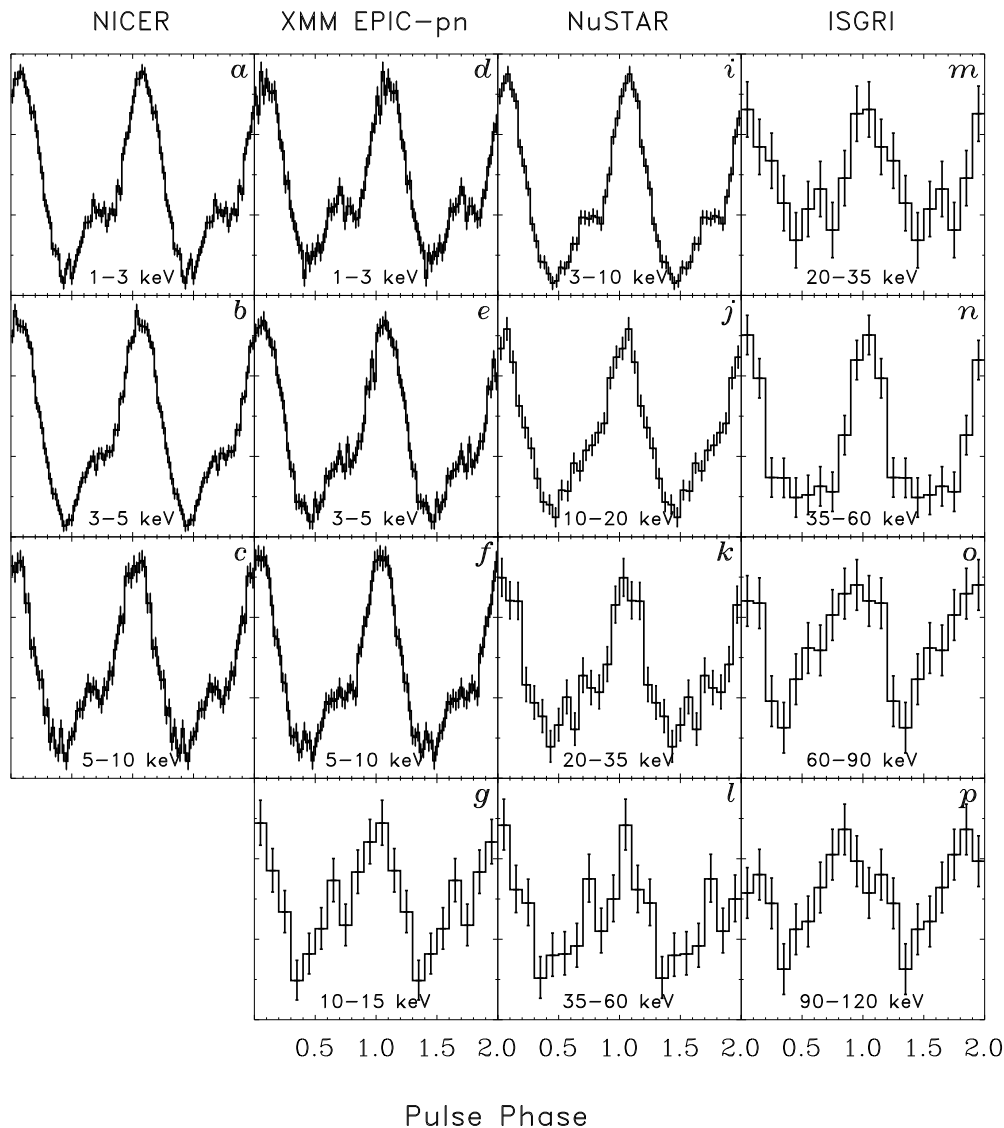}
  \caption{High-energy pulse profiles as function of energy for \psrtar\ using NICER (MJD 58345 -- 58364; segment-1), XMM-Newton, NuSTAR (MJD 58344; ToO-I) and INTEGRAL ISGRI (Revs. 1986-2006) data
           covering the 1 -- 120 keV energy band. Pulsed emission has been detected significantly up to $\sim 120$ keV.
 }
  \label{pp_collage}
\end{figure*}

We could identify two time intervals in the \nicer\ observation sequence for which the timing behaviour of \psrtar\ was relatively stable.
The first segment lasted for about 20 days from MJD 58345 -- 58364 (segment-1), and the second one for about 14 days from MJD 58370 -- 58383 (segment-3). Between (segment-2) and beyond these segments the pulsar showed noisy/erratic behaviour, very likely due to variations in the accretion torques. The best fit parameters for $\nu$ and $\dot\nu$
for these two segments are shown in Table \ref{table:eph} along with the epoch $t_0$, the root-mean-square, RMS, of the fit and $\Phi_0$, a phase offset introduced to keep consistent alignment
between different ephemerides. 

For the first segment we measure a significant {\it positive\/} value for $\dot\nu$, indicating that spin-up is active during this period, while for segment-3 a constant spin frequency only 
adequately fitted the data. Between these segments, in segment-2 (MJD 58364 -- 58370), a small drop in frequency of $\Delta\nu \simeq -1.6\cdot 10^{-7}$ Hz occurred. Note, that the \xmm\ ToO observation was performed at the start of this period. The full set of \nicer\ pulse arrival time residuals with respect to the timing model of segment-3 (i.e. model with {\em constant\/} frequency) is shown in Fig. \ref{toa_residuals}.

It is clear from this plot that there is no any correlation between the residuals and the flux state, as represented by the \nicer\ 1--10 keV background-subtracted count rates (magneta points), of \psrtar. Note, that an equivalent X-ray (\nicer\ and \swift-XRT) lightcurve profile is given in \citet{Gusinskaia2020}, who performed a detailed X-ray / radio comparison.

We also investigated the impact of the location inaccuracy of $0\farcs03$ on the derived frequency and frequency derivative using the formulae given in appendix A of \citet{hartman2008}. The frequency and frequency derivative offsets are: $\Delta\nu \la 8$ nHz and $\Delta\dot{\nu} \la 1.6\times 10^{-15}$ Hz\,s$^{-1}$, respectively. The first value is comparable to the uncertainty in the measured frequency, while the second one
is much smaller than the quoted frequency derivative uncertainty for segment-1, and so the location inaccuracy has no significant impact on the reconstructed values for the frequency and frequency derivative.

\subsection{High-energy pulse profiles: \nicer, \xmm, \nustar\ and \Integ-ISGRI}
\label{sect_tm_pp}
The availability of accurate phase-coherent ephemerides (see Table \ref{table:eph}) made a (multi-instrument) detail study possible of the energy dependence of the pulse profile of \psrtar.
Pulse-phase folding the \nicer\ barycentered event times, corrected for the orbital motion, from segment-1 (MJD 58345 -- 58364; exposure 49.61 ks) observations, according to 
\begin{equation}
\Phi(t)=\nu\cdot (t-t_0) + \frac{1}{2}\dot\nu\cdot (t-t_0)^2 - \Phi_0 \label{eq:folding}
\end{equation}  
yielded the pulse-phase distributions shown in Fig. \ref{pp_nicer_vs_e} for three different measured energy bands: 1--3 (black), 3--5 (orange) and 5--10 keV (red).

\begin{figure*}[t]
  \centering
  \includegraphics[angle=0,width=8.0cm]{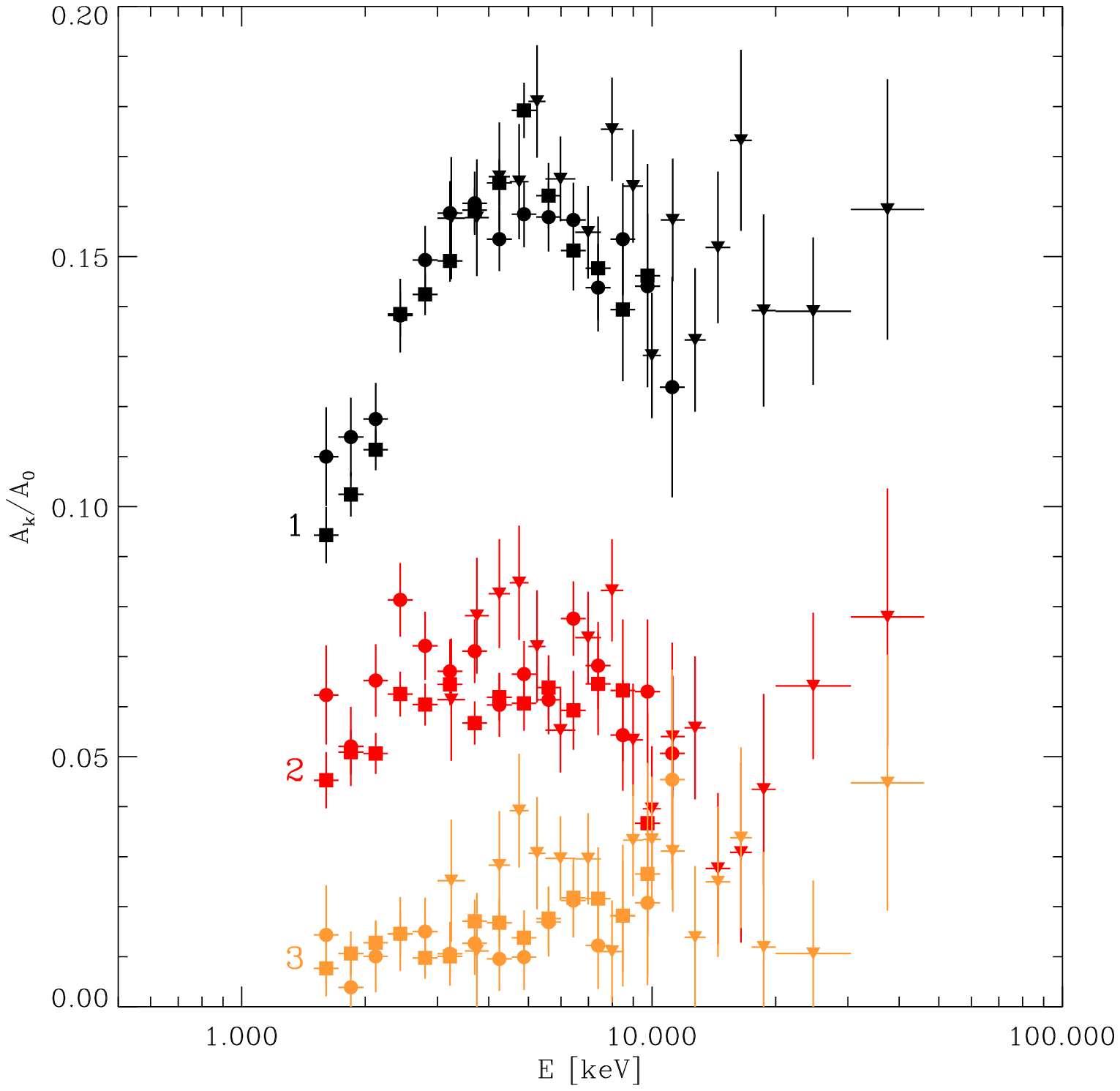}
  \includegraphics[angle=0,width=8.0cm]{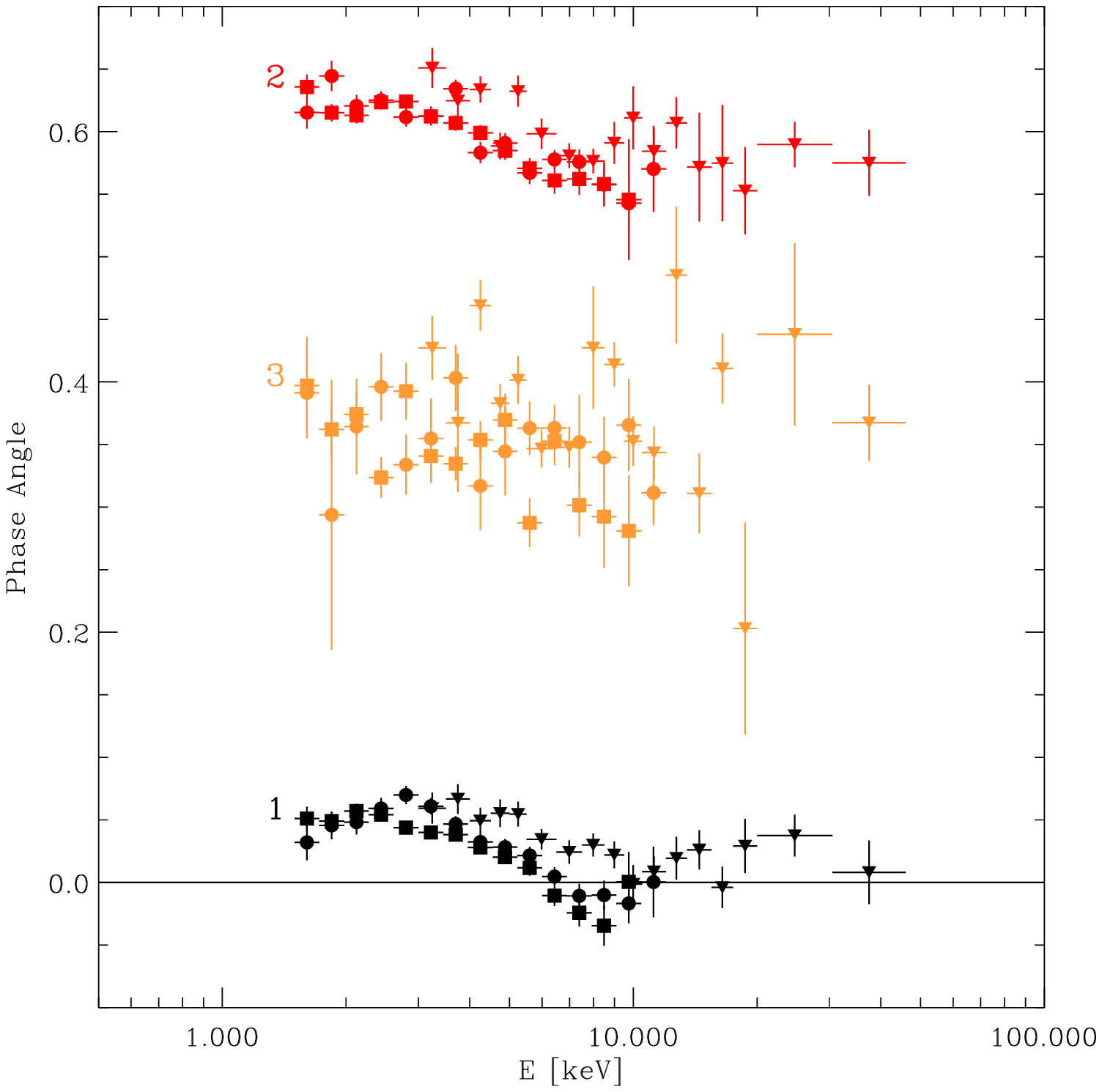}
  \caption{(Left) The fractional amplitude of various harmonics (1 - fundamental, 2 - first overtone, and 3- second overtone) as a function of energy using background 
           corrected data from \xmm\ EPIC-pn (circles), NICER (squares) and \nustar\ (triangles).
           The pulsed fraction of the fundamental component increases from $\sim$ 10\% till $\sim 17$\% from 1.5 to 3--4 keV, from where it more or less 
           saturates at a level of $\sim 17$\%.
           (Right) The phase-angle as function of energy for the three Fourier components. The shift towards lower phases from $\sim 1.5$ keV till $\sim 10$ keV for both the fundamental and first overtone is evident.
          }
  \label{pp_fourier_fit}
\end{figure*}

From this (\nicer\ only) plot it is already clear visually that the more energetic photons of the double-peaked pulsed emission component arrive earlier than the softer photons. In the next 
section a detailed profile analysis including \nicer, \xmm\ and \nustar\ data will provide all quantitative information related to this across the $\sim$1--50 keV band.

We also studied the stability of the pulse shape in the 1--3 keV \nicer\ band as a function of time, and found no significant changes, i.e. the profile morphology was stable across the time window that \nicer\ observed \psrtar\ in spite drastic flux changes.

The \nicer\ 1--3 keV data at various outburst stages (see Fig. \ref{toa_residuals}) were further used to determine the relation between the  pulsed flux $C_{\rm Pulsed}$ and (background corrected) total flux $C_{\rm total}$. We found that both quantities were tightly correlated, as expected, with a $C_{\rm Pulsed}/ C_{\rm total}$ ratio of $\sim 11\%$ consistent with the results shown later in Sect. \ref{sect_tm_rsl} for the 1--3 keV band, and so the pulsed fraction is independent on the flux state of the source.

Next, we applied the timing models shown in Table \ref{table:eph} in timing analyses of the other high-energy instruments adopting methods, screening procedures and selections outlined in the instrument Sects.
\ref{instr_integral}, \ref{instr_nustar} and \ref{instr_xmm}, for \Integ-ISGRI, \nustar\ and \xmm\ EPIC-Pn, respectively.

The \xmm\ data were taken at the very beginning of segment-2 (for which no phase-coherent ephemeris is available) very near the end of segment-1, and so the timing data were folded upon the 
parameters listed under ephemeris-1 in Table \ref{table:eph} realizing that a small phase shift can occur between the \nicer, as baseline, and \xmm\ EPIC-Pn profiles. Cross-correlation of
\xmm\ and \nicer\ 1--10 keV pulse-profiles did indeed show a small phase shift of $-0.043$ as expected. Even above 10 keV a significant pulsed signal was detected at a $6.2\sigma$ confidence level
adopting a $Z_2^2$-test (see e.g. panel g in Fig. \ref{pp_collage}).

The \nustar\ data has been taken from ToO-I which has the highest statistics among the two \nustar\ ToO's. Application of the correction method reported by \citet{gotthelf2017}, using ephemeris-1
parameters and the \nicer\ 3--10 keV pulse-profile as phase-alignment baseline yielded pulsed signal detections up to $\sim$50--60 keV:  the 35--60 keV band yielded a $4.6\sigma$ pulsed signal detection 
applying a $Z_2^2$-test (see e.g. panel l in Fig. \ref{pp_collage}). 

\Integ-ISGRI timing data from Revs 1986, 1989, 1992, 1994--1995, 1998, 2001--2003 (and 2004--2006) have been folded upon the ephemerides listed in Table \ref{table:eph}, some in extra-polation mode.
Pulsed emission has been detected significantly up to $\sim 120$ keV with a signal strength of about $4.1\sigma$ in the 90--120 keV energy band (see e.g. panel p in Fig. \ref{pp_collage}).

The full set of pulse-phase distributions resulting from pulse-phase folding the barycentered event times upon the ephemeri(de)s parameters listed in Table \ref{table:eph} is shown in Fig. \ref{pp_collage} 
for the four different high-energy instruments covering the 1--120 keV band: \nicer\ (1--10 keV; left panels a-c), \xmm\ (1--15 keV; middle left panels d-g), \nustar\ (3--60 keV; middle right panels i-l) and
\Integ\, ISGRI (20--120 keV; right panels m-p).

\subsection{Timing results: pulsed fraction, phase-angle}
\label{sect_tm_rsl}
To obtain quantitative information about morphology changes of the pulse-profile as a function of energy we produced pulse-phase distributions in narrow energy bands for \nicer, \xmm\ and \nustar\ covering the $\sim$ 1--50 keV band (all segment-1 data except \xmm\ (early segment-2; see Sect. \ref{sect_tm_sol}), folded upon segment-1 ephemeris).   These measured distributions ${\cal N}(\phi)$ were subsequently fitted with a truncated Fourier series ${\cal F}(\phi)$ given by 
\begin{equation}
{\cal F}(\phi)= A_0 + \sum_{k=1}^n a_k \cos(k\phi) + b_k \sin(k\phi) = A_0 + \sum_{k=1}^n A_k \cdot \cos(k\cdot (\phi-\phi_k)) \label{eq:fourfit}
\end{equation}
with $\phi_k=\arctan({b_k}/{a_k})$ and $A_k=\sqrt{a_k^2+b_k^2}$. For each harmonic maxima can be found at $\phi_{\max}=\phi_k \bmod (2\pi/k)$ (in radians).

The results of these fits are shown in Fig. \ref{pp_fourier_fit} with the fractional background corrected amplitude $A_k/A_0$ in the left panel, and the phase angle
$\phi_k$ in the right panel, now mapped from radians to (pulse) phase i.e. range [0,1]. The fundamental, $k=1$, is colored black, while the first, $k=2$, and second, $k=3$,
overtone are colored red and orange, respectively.
It is clear from these energy-resolved results that the pulsed fraction of the fundamental increases from about 10\% to 17\% moving from 1.5 to 3--4 keV, and beyond it fluctuates a bit around 17\%.
This is consistent with the energy integrated result reported by \citet{Sanna2018}. The first overtone shows a $\sim$ 6\% contribution and the second a $\sim$ 1--2\% to the pulsed fraction, also
consistent with \citet{Sanna2018}.

From the phase angle plot (Fig. \ref{pp_fourier_fit} right panel) the decreasing trend in angle for increasing energies till $\sim$ 10 keV for both the fundamental and first overtone is evident,
meaning that the higher energy photons of the pulsed component arrives earlier than the softer ones. This was already shown for \nicer\ data alone in Fig. \ref{pp_nicer_vs_e}. The total phase 
decrease is about 0.05 ($\sim 95\,\mu$s) going from 1.5 to 10 keV.


\section{Broad-band spectral analysis of the total emission} 
\label{sect_spc}
In this section, we report on the analysis of the averaged broad-band spectral properties of \psrtar, since this source was intensively observed with different observatories covering
a broad energy range. The different pointings were taken over the whole outburst, however, covering different intensity states of the source (see Fig.~\ref{outburst_profile}). 
Therefore, to produce the broad-band energy spectrum we used \xmm, \nustar\ and \Integ\ observations, performed either close in time or during similar intensity states. For the 
high-energy part we make use of the \Integ\ ISGRI spectrum covering the 30 -- 150 keV energy range, as obtained during revolutions 1994 and 1995, which covers the \xmm\ observation 
(see Fig. \ref{outburst_profile}).  We fit these data together with the \xmm\ EPIC-pn (1.5--12 keV)\footnote{We ignored events with energies smaller than 1.5 keV, because from the \xmm\ EPIC-pn timing analysis we concluded that the redistribution in the {\sc rmf} model underestimates the (unexpected, given the high $N_{\rm H}$ in the range (2--5) $\times 10^{22} $\,cm$^{-2}$) measured pulsed signal of $3.2\sigma$ in the 0.3--0.5 keV band.}, \xmm\ EPIC-MOS2 (0.5 -- 10 keV), RGS 1 and 2 ($\sim0.33$ -- 2.1 keV) data from
\xmm\ observation 0795750101, and \nustar\ FPMA/FPMB data (3.5 -- 79 keV) from observation 90401331002. All \xmm\ EPIC-pn and EPIC-MOS2 and \nustar\ spectra of the source were optimally rebinned using the prescription in paragraph 5.4 of \citet{Kaastra2016}. The \xmm\ RGS spectra of the first dispersion order were rebinned with a minimum number of 25 photons per energy bin. The spectral analysis was carried out using {\sc xspec} version 12.6 \citep{arnaud96}. All uncertainties in the spectral parameters are given at a $1\sigma$ c.l. for a single parameter.  We applied an additional systematic uncertainty of 2 per cent for each instrument due to uncertainties in the spectral responses of each involved instrument.

\begin{figure*}[t]
  \centering
  \includegraphics[angle=0,width=7.5cm,height=6.75cm,bb=35 270 580 710,clip=]{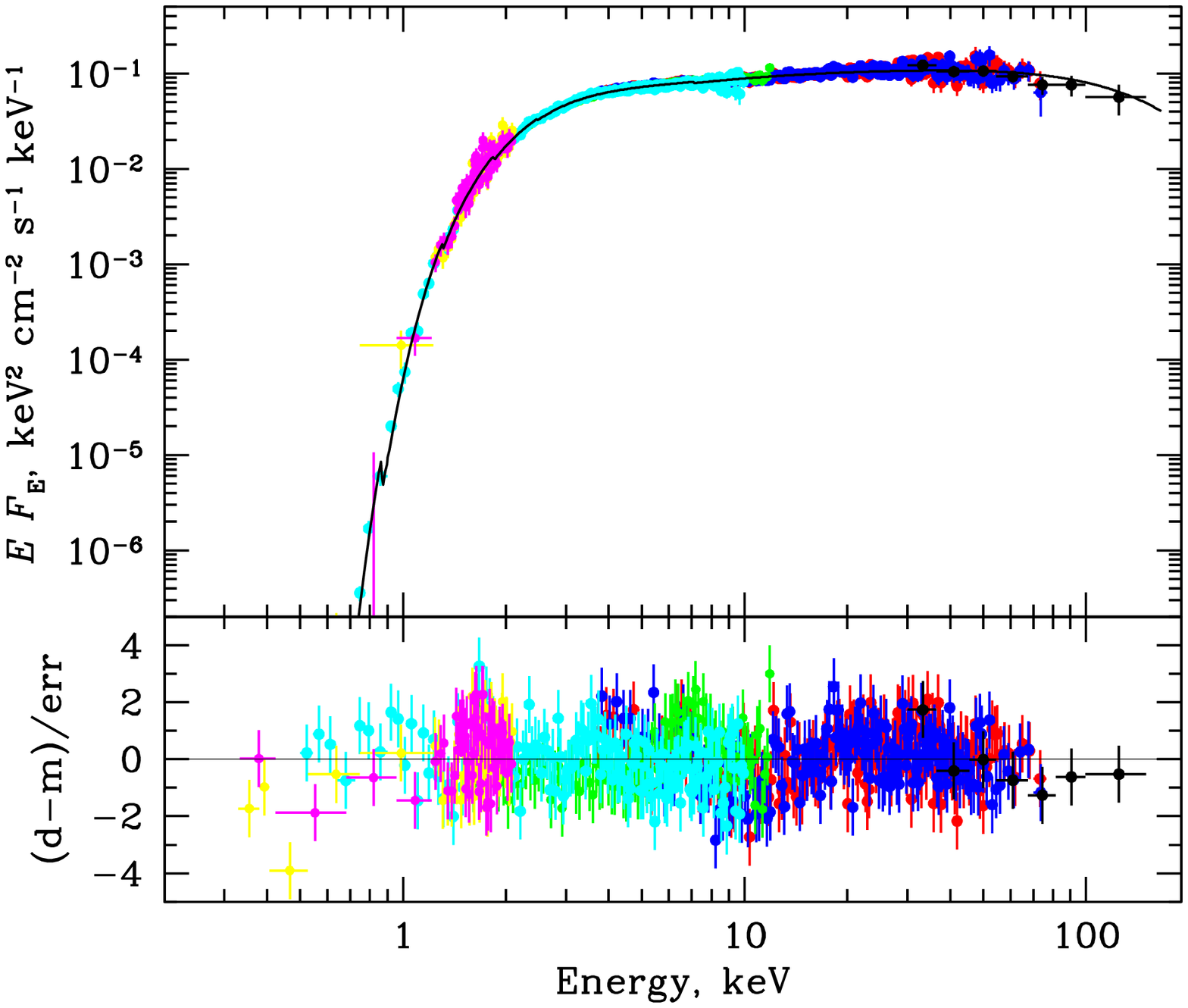}
  \includegraphics[angle=0,width=8cm,height=8cm]{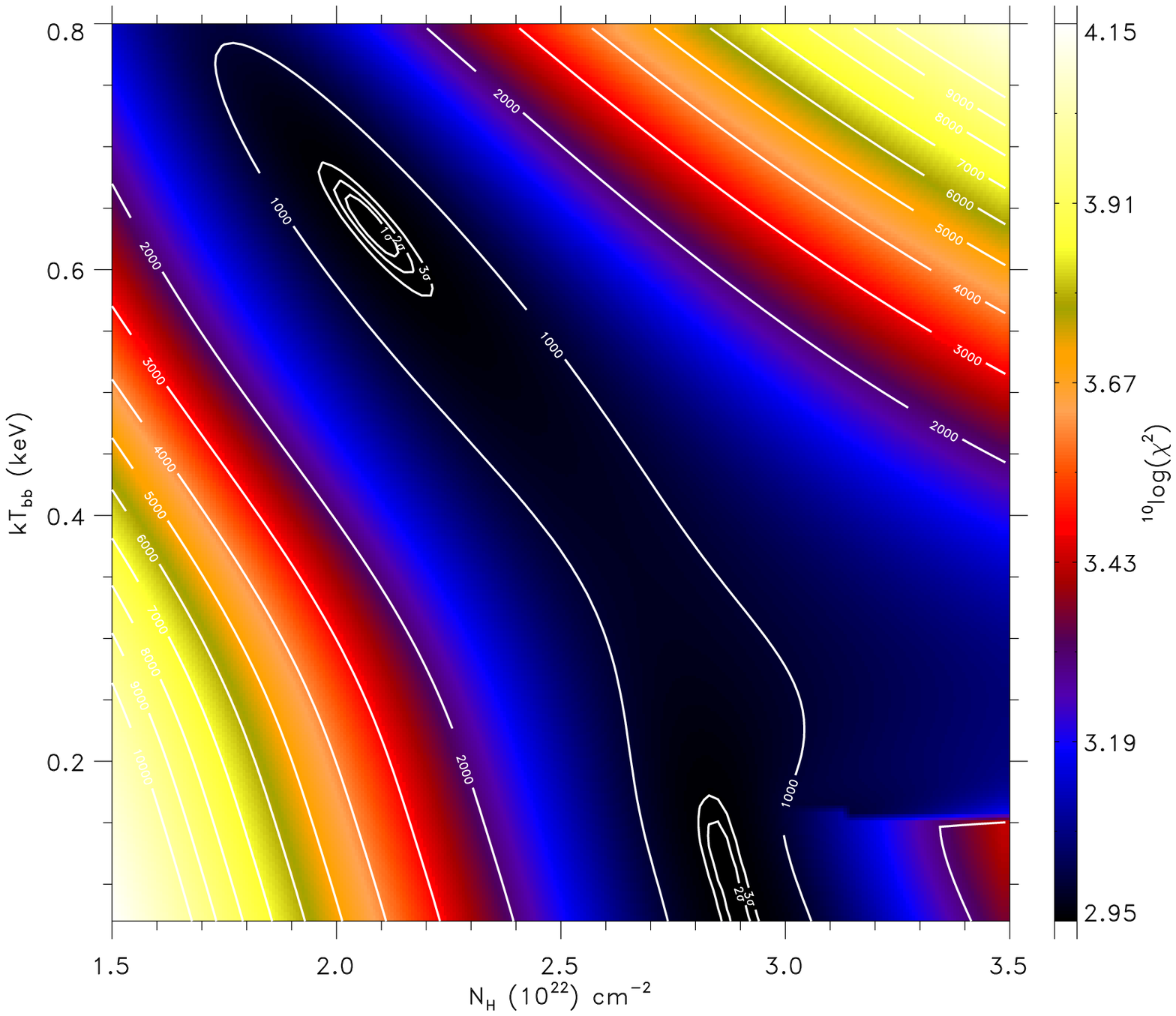}
  \caption{Unfolded absorbed broad-band spectrum of \psrtar\ in the 0.3--150 keV energy range (left panel). The data points are obtained from the two \xmm\ RGS instruments (yellow and magenta datapoints, $\sim0.33$--2.1 keV), \xmm\ EPIC-pn (green points, 1.5--12 keV), \xmm\ EPIC-MOS2 (cyan  points, 0.5--10 keV), \nustar\ FPMA/FPMB (red and blue points, 3.5--80 keV), and \Integ-ISGRI (black points, 30--150 keV). The fit is obtained with the \compps\ model, represented in the top panel with a solid black line. The residuals from the best fit are shown at the bottom. The right hand panel shows the correlation between $N_{\rm H}$ and $kT_{\rm{BB}}$ in the \compps\ model: two minima co-exist. The contours encompassing the minima denote the 1, 2 and 3$\sigma$ confidence levels assuming 2 d.o.f., while the others are labeled with their corresponding $\chi^2$-value (note: d.o.f. is 910).
          }
  \label{compps_total_spectrum}
\end{figure*}

At first, we fitted the combined spectra using a simple absorbed power-law model, including also normalization constants in the fit to take into account uncertainties in the cross-calibrations of the instruments and the source variability (the data are not covering strictly the same time interval of the source outburst). This first fit provided an absorption column density of $N_{\rm H}=(3.01\pm0.02) \times 10^{22} $\,cm$^{-2}$, a power-law photon index $\Gamma=1.84\pm0.01$ and a relatively large $\chi^{2}_{\rm red}{\rm /d.o.f.} = 1.15/912$, due to the `wavy' residuals at all energies and significant deviations from a pure power law at energies above $\sim80$~keV. Although the fit was not formally acceptable, we noticed that the values of the normalization constants were all in the range $1.0\pm0.2$, compatible with the finding that the average spectral properties of the source during the considered part of the outburst were stable. A similar range for the normalization constants was also found in all other fits reported below (we assumed in all cases the \nustar\ FPMA as our reference instrument and fixed its normalization constant to unity).
Our derived $N_{\rm H}$ value is inconsistent with the value of $(4.4\pm0.3) \times 10^{22} $\,cm$^{-2}$ reported by \citet{Russell2018}, who used in an absorbed power-law fit only \swift-XRT data covering the 0.5--10 keV band, and so lacking spectral band-width and sensitivity.

We also fitted the averaged broad-band spectrum of \psrtar\ with an absorbed thermal Comptonization model \nthcomp\ in {\sc xspec} \citep[see e.g.][]{zdziarski1996,zycki1999} 
to take into account the emission produced by a thermal distribution of electrons which Compton up-scatter the soft seed X-ray photons.   
Our fit has 5 free fit parameters: the absorbing hydrogen column density $N_{\rm H}$, seed photon temperature $kT_{\rm bb, seed}$, electron temperature $kT_{\rm e}$, asymptotic power-law photon-index $\Gamma$ and normalization factor $K$, while we fixed the input type for the seed photons to 0 (i.e. blackbody seed photons) and put the redshift $z$ to 0 
for this model.
This more physically motivated spectral model provided a statistically better fit than the phenomenological power-law model described above. Although the results of this fit, summarized in Table~\ref{table:spec}, are quantitatively similar to those previously reported, we noticed that our measured absorption column density $N_{\rm H}$ of $(2.25\pm0.05) \times 10^{22} $\,cm$^{-2}$, is considerably lower than the $N_{\rm H}$ value of $(3.6\pm1.1) \times 10^{22} $\,cm$^{-2}$ reported by \citet{Sanna2018} who adopted an equivalent spectral model across the 0.5--80 keV band using \swift-XRT (0.5--7.5 keV), \nustar\ 3--70 keV), \nicer\ ($> 1.4$ keV) and \Integ-ISGRI (30--80 keV). We ascribe the difference to the fact that we are using a wider energy range, especially the coverage below $\sim1.5$ keV by including accurate spectral data from the high-sensitivity instruments \xmm\ RGS 1 \& 2 and EPIC-MOS2 (compared to \swift-XRT) down to 
$\sim0.33$ keV. Our value of $N_{\rm H}$ is compatible with that from model C reported by \citet{Nowak2019} (providing the `fairest estimate for the equivalent neutral column'; see their Sect. 4.2), $N_{\rm H}=(2.9\pm0.5) \times 10^{22} $\,cm$^{-2}$, obtained using \cxo-HETG 1--9 keV data and also assuming a Comptonized blackbody continuum. The main difference between the results of \citet{Sanna2018} and \citet{Nowak2019} is the temperature of the seed photons, $kT_{\rm bb, seed}\sim 0.79$ keV and $kT_{\rm bb, seed}\sim 0.06$ keV, respectively. Our best-fit value for the blackbody seed photons, however, is $kT_{\rm bb, seed}\sim 0.52$ keV (see Table \ref{table:spec}).

To compare the total broad-band emission spectrum of \psrtar\ with those of other AMXPs observed at hard X-rays with \Integ\ \citep[e.g.,][]{gdb02,gp05,mfa05,mfb05,mfc07,ip09,falanga11,falanga12,deFalcoa,deFalcob,ZLI2018}, we performed also a spectral fit using a thermal Comptonization model (\compps) in the slab geometry \citep{ps96}. The main fit parameters are the absorption column density $N_{\rm H}$, the Thomson optical depth $\tau_{\rm T}$ across the slab, the electron temperature $kT_{\rm e}$, the temperature $kT_{\rm bb, seed}$ of the soft-seed thermal photons (assumed to be injected from the bottom of the slab), and
the normalization of the model $K$, which for black body seed photons equals $K=(R_{\rm km}/D_{10})^2$ with $R_{\rm km}$ the radius of the source in km and $D_{10}$ the source distance in units 10 kpc. The results of this fit are reported in Table~\ref{table:spec} and are compatible with those measured from other AMXPs in outburst. We show in left panel of Fig.~\ref{compps_total_spectrum} the absorbed (unfolded/deconvolved) total broad-band emission spectrum of \psrtar,\ together with the residuals from the best-fit model at the bottom. 

\begin{table}[h] 
{\small
\caption{Best parameters determined from the fits to the total broad-band emission spectrum 
of \psrtar\ performed with the \xmm, \nustar\ and \Integ\ using the {\sc phabs} absorbed comptonization models \nthcomp\ or \compps, respectively.}
\centering
\begin{tabular}{lccc} 
\hline 
Parameter & Unit & \nthcomp\tablefootmark{a} & \compps\tablefootmark{a} \\
\hline 
\noalign{\smallskip}  
$N_{\rm H}$                  & $(10^{22} {\rm cm}^{-2})$ & $2.25\pm0.05$ & $2.09\pm0.05$\\ 
$kT_{\rm bb, seed}$          &(keV)                      & $0.52\pm0.02$ & $0.64\pm0.02$\\ 
$\Gamma$                     &                           & $1.815\pm0.005$ & --           \\
$kT_{\rm e}$                 &(keV)                      & $27^{+3}_{-2}$  & $38.8\pm1.2  $\\ 
$\tau_{\rm T}$               &                           & --            & $1.59\pm0.04$\\ 
$K$                          &(km$^2$)                   & --            & $223\pm24 $\\
$\chi^{2}_{\rm red}/{\rm d.o.f.}$  &  & 1.05/910      & 1.08/910 \\
$F_{\rm bol}$ \tablefootmark{b}    & ($10^{-10}$erg cm$^{-2}$ s$^{-1}$) & $6.30\pm0.02$ & $6.40\pm0.02$\\
\noalign{\smallskip}  
\hline  
\end{tabular}  
\tablefoot{ \tablefoottext{a}{Uncertainties are given at $1\sigma$ confidence level.}
            \tablefoottext{b}{Unabsorbed flux in the 0.3 -- 150 keV energy range.}
          }
\label{table:spec} 
}
\end{table} 

The absorption column density measured from the fit with the \compps\ model is a bit less, but compatible at the $2\sigma$-level with the value obtained using the \nthcomp\ model. It is noteworthy that our derived $N_{\rm H}$ of $\sim2.09 \times 10^{22}$\,cm$^{-2}$ is consistent with the estimate of the {\it total\/} Galactic column contribution of $\sim2.2 \times 10^{22}$\,cm$^{-2}$ based on reddening maps \citep[see Sect. 3.3 of][for a discussion on this]{Russell2018}, and thus there is no need for additional intrinsic absorption from the source environment. 
Our value is somewhat larger than the {\it atomic\/} neutral hydrogen column $N_{\rm H, Gal}$ of (1.12--1.44)$\times 10^{22}$\,cm$^{-2}$ \citep{Dickey1990,Kaberla2005}, however, the difference can be attributed
to a {\it molecular\/} hydrogen contribution, which can be quite significant in the Galactic bulge direction.

We studied the correlation between the fit parameters in the \compps\ model. In particular, the 2d-{\sc xspec} fitting result for the correlation between $N_{\rm H}$ and $kT_{\rm bb, seed}$ has two local minima corresponding to two pairs of $N_{\rm H}$ and $kT_{\rm bb, seed}$ values. The main minimum corresponds to a low absorption column $N_{\rm H} \sim 2.09 \times 10^{22}$\,cm$^{-2}$ and a high seed photon temperature $kT_{\rm bb, seed}\sim 0.64$~keV, whereas the other minimum gives a higher $N_{\rm H}$ of $\sim 2.93 \times 10^{22}$\,cm$^{-2}$ and a lower seed photon temperature $kT_{\rm bb, seed}\sim 0.05$~keV, the latter seems to be compatible with spectral model C given in \citet{Nowak2019}. The existence of the two minima might also explain the large difference in the blackbody temperatures obtained by \cite{Sanna2018} and \cite{Nowak2019}. This situation is illustrated in the right hand panel of Fig. \ref{compps_total_spectrum} where the $1,2$ and $3\sigma$ confidence, assuming 2 degrees of freedom, contours are shown around the (two) minima for the $N_{\rm H}$ versus $kT_{\rm bb, seed}$ correlation as well as contours labeled with their corresponding $\chi^2$-values, starting at 1000 in steps of 1000 (note d.o.f. is 910). A similar (more mildly) correlation exists between the \compps\ model parameters $kT_{\rm e}$ and $\tau_{\rm T}$.

To assess possible (reflection related) emission from the iron K$_\alpha$ complex \citep[near 6.40 keV; see e.g.][]{Sanna2018} we added a Gaussian line with free energy location, width and normalization to the absorbed
\compps-model. Though the fit formally improves at a $4.9\sigma$ level (for 3 parameters) the `best' line location of $6.70_{-0.15}^{+0.14}$ keV and line width of $0.69_{-0.15}^{+0.18}$ keV are too far off and too broad, respectively. The emission feature could be a manifestation of a blend of lines from different Fe ionisation stages when measured with CCD spectral resolution. 
However, more likely it reflects uncertainties in the \xmm\ EPIC-pn (in timing mode) response description (see the green data points in the bottom panel of Fig. \ref{compps_total_spectrum}), because in the \cxo-HETG data with superior energy resolution (with respect to the EPIC-Pn CCD spectral resolution) there is no evidence for iron K$_\alpha$ emission \citep[see Fig. 6 of][]{Nowak2019}, and so we do not consider this as a real detection.

An advantage of the \compps\ model is that it allows us to estimate the apparent radius $R_{\rm km}$ of the thermally emitting region on/near the NS surface from $R_{\rm km}=\sqrt{K}\cdot D_{10}$. At the distance of \psrtar\ of $d\sim 7.6$~kpc (see Sect. \ref{sec:burstprop}), the radius of this region is $11.3\pm0.5$~km, compatible with 
the radius of a neutron star.

\section{A thermonuclear (type-I) burst detected in \Integ\ JEM-X data}
\label{sec:burst}
Thermonuclear (type-I) X-ray bursts are produced by unstable burning of accreted matter on the NS surface. The spectrum from a few keV to higher energies can usually be well described  as a blackbody with temperature $kT_{\rm bb}\sim$1--3 keV. The energy-dependent decay time of these bursts is attributed to the cooling of the NS photosphere and results in a gradual softening of the burst spectrum \citep[see][for a review]{lewin93,Strohmayer06}. 

We scrutinized the \nicer, \xmm\ EPIC-MOS/PN, \swift-XRT and \nustar\ data for the presence of thermonuclear bursts at different time scales (1, 10 and 100~s), but we found none. Next, we focused on \Integ\ JEM-X (3--35 keV) data, and identified one thermonuclear burst candidate. Below we describe the analysis of the JEM-X data covering the event.

\subsection{Properties of the type-I X-ray burst} 
\label{sec:burstprop}
We searched \Integ/JEM-X data for bursts within revolutions 1986--2009, following the same approach as \citet{Mereminskiy2017}. 
We focused on one of the two JEM-X cameras, JEM-X1;  JEM-X2 was operating simultaneously, but this instrument seems to have reduced sensitivity in recent years.
We identified one  candidate thermonuclear (type-I) burst from \psrtar, in Scw-500010 of \Integ\ revolution 2001 on Sept. 19, 2018; 23:07:33 
(MJD 58380.96358~TT).
The key properties of the burst are listed in Table \ref{tab:burstprop}.
Based on its intensity, this burst should have also been detectable by ISGRI; however, due to the loss of sensitivity in the past years in the lower ISGRI energy band, 18--27~keV, we did not detect this event in those data.

\begin{figure}[t] 
\centering
 \includegraphics[angle=-90,width=7.5cm]{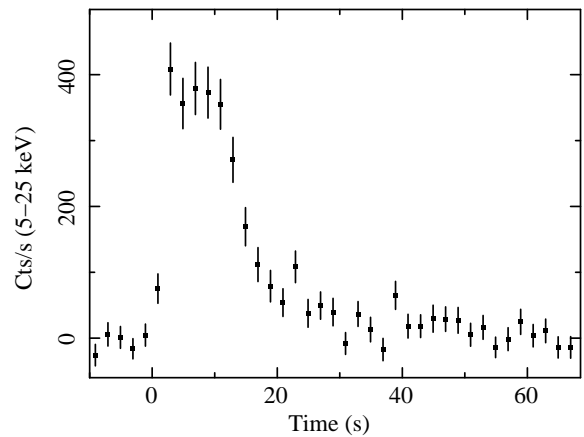}
\caption{The light curve of the type-I X-ray burst detected by JEM-X and reported for the first time in this paper. The burst start time was 58380.96358~MJD. The JEM-X light curve 
was extracted in the 5--25 keV energy range with a time bin of 2~s.}
\label{fig:burst}
\end{figure}

The type-I X-ray burst occurred $\sim 60$~d after the outburst commenced, just after the source was at the highest X-ray luminosity (see Fig. \ref{outburst_profile}). We defined the burst start time as the time at which the X-ray intensity of the source first exceeded 10\% of the burst peak flux (above the persistent intensity level). The event showed a fast ($\sim1$--2~s) rise time, followed by a $\sim 10$~s long plateau at its peak (Fig. \ref{fig:burst}). Therefore, we conclude that most likely a photospheric radius expansion took place \citep[see e.g.,][]{mfc07,galloway08b,Bozzo10}. After the plateau, the burst decay profile could be well fitted with an exponential function and the correspondingly derived e-folding time is $\tau_{\rm fit}\sim9$ s. The total duration of the burst was  $\sim 40$ s.

For the burst spectral analysis, we first determined the  persistent flux before the burst onset by converting the ISGRI counts from Rev-2001 to the total broadband spectrum obtained from Revs. 1994--1995 (see Sect. \ref{sect_spc}). We note, that during the outburst the ISGRI data in the 20--150 keV band alone were fitted by a power-law model, and the spectral parameters were stable. The pre-burst flux was $F_{\rm pers,bol}\sim 1.2\times10^{-9}$~erg~cm$^{-2}$ s$^{-1}$.

We analysed the burst spectra over the 3--25 keV band, and afterwards extrapolated to the 0.1--40 keV broad-band. This is justifiable for the JEM-X data since the blackbody temperature is well inside the spectral energy coverage. We carried out two spectral analyses of the JEM-X data covering the burst event, one to determine the burst peak flux during the 10~s (plateau) time interval, and a second to determine the fluence integrated over the total burst duration of 40~s.

The net burst spectra were both well-fitted ($\chi^{2}_{\rm red}$=0.98) adopting a simple {\texttt tbabs*bb} model, with absorption column fixed at $N_{\rm H}$=2$\times$10$^{22}$~cm$^{-2}$.  The estimated fluence was $f_{\rm b}\sim 1.1\times 10^{-6}$ erg cm$^{-2}$ and peak flux $F_{\rm peak} \sim 7.6\times 10^{-8}$ erg cm$^{-2}$s$^{-1}$.  

We estimated the burst energy including consideration of the expected anisotropic X-ray emission arising from the non-axisymmetric distribution of mass around the neutron star. The total X-ray luminosity $L_{\rm X}$ is estimated from the flux $f_{\rm X}$ as
\begin{equation}
    L_{\rm X\,b,p} =  4\pi d^2 f_{\rm X\,b,p}\, \xi_{\rm b,p},
\end{equation}
where $d$ is the distance to the source, $\xi_{\rm b,p}$ is the anisotropy factor, and the subscripts b and p correspond to the burst and persistent emission, respectively \citep{fuji88}.
\citet{he16} carried out simulations to predict the anisotropy factor based on different disk geometries; here we adopt their model ``a'', corresponding to a flat, optically thick disk. Over the relatively narrow inclination range of 28--30$^\circ$, the predicted anisotropy factor for the burst (persistent) flux is $\xi_{\rm b} = 0.71$ ($\xi_{\rm p} = 0.49$). Values $\xi_{\rm b,p}<1$ imply the flux is preferentially beamed toward us, so that the inferred isotropic luminosity would be significantly above the actual value. We estimated the distance by equating the peak flux with the empirical Eddington luminosity of 
\citet{kuulkers03} and including the anisotropy factor, as
$7.6_{-0.6}^{+0.8}$~kpc.
The corresponding bolometric persistent luminosity prior to the burst was $4.1\times10^{36}$~erg~s$^{-1}$, (i.e. $\sim1.1$\%~$L_{\rm Edd}$, where $L_{\rm Edd} = 3.8\times10^{38}$~erg~s$^{-1}$ is the Eddington luminosity).

\begin{table}[h] 
\caption{Parameters of the type-I X-ray burst observed by \Integ\ JEM-X during the outburst of \psrtar\ in 2018.
\label{tab:burstprop}}
\centering
\begin{tabular}{ll} 
\hline 
\hline 
$T_{\rm start}$\tablefootmark{a} (UTC)& 2018-09-19 23:07:33\\
$\Delta t_{\rm burst}$ (s)  & $40\pm2$\\
$\Delta t_{\rm rise}$ (s) & $2\pm1$\\
$\tau_{\rm fit}$\tablefootmark{b} (s) & $8.5\pm1.0$\\
$F_{\rm peak}$\tablefootmark{c}($10^{-8}$ erg cm$^{-2}$ s$^{-1}$) & $7.6\pm1.4$\\
$kT_{\rm bb,peak}$\tablefootmark{d} (keV) & 2.55$\pm0.16$\\ 
$F_{\rm pers, bol}$\tablefootmark{e} ($10^{-9}$ erg cm$^{-2}$ s$^{-1}$) & $1.2\pm0.2$\\
$f_{\rm b}^*$\tablefootmark{f} ($10^{-6}$ erg cm$^{-2}$) & $1.1\pm0.1$\\
\hline  
\end{tabular}  
\tablefoot{ 
\tablefoottext{a}{58380.96358 MJD.}
\tablefoottext{b}{The $\tau_{\rm fit}$ has been measured after a plateau of $\sim$10 s.}
\tablefoottext{c}{Burst peak flux in 0.1--40 keV energy band.}
\tablefoottext{d}{Burst peak temperature.}
\tablefoottext{e}{Pre-burst unabsorbed flux in 0.1--250 keV energy range.}
\tablefoottext{f}{Burst fluence in 0.1--40 keV energy band.}
}
\label{tab:burst} 
\end{table} 

\subsection{Inferred burst ignition conditions} 
\label{sec:igncond}

To constrain the ignition conditions that gave rise to the JEM-X burst, we first determine the local accretion rate per unit area onto the compact object at the time of the event as $\dot{m} = L_{\rm pers} (1+z) (4\pi R^2(GM/R_{\rm NS}))^{-1}$, i.e., 
$\dot m \sim 2\times10^3$ g cm$^{-2}$ s$^{-1}$ (where the gravitational redshift is $1+ z = 1.259$ for a canonical NS with a mass $M_{\rm NS}=1.4M_\odot$ and a radius of $R_{\rm NS}=11.2$ km).
We can then estimate the ignition depth at the onset of the burst as 
\begin{equation}
y_{\rm ign} = 
f_{\rm b}(1+z)(4\pi R^2Q_{\rm nuc})^{-1}   \label{eq:yign} 
\end{equation}
where the nuclear energy 
$Q_{\rm nuc}$ depends on the mean hydrogen fraction at ignition $\langle X\rangle$, as $Q_{\rm nuc}\sim 1.35+6.05\langle X\rangle\ {\rm MeV/nucleon}$\footnote{This expression is an update to one previously used in the literature,
$Q_{\rm nuc}\sim 1.6+4\langle X\rangle\ {\rm MeV/nucleon}$ \citep[e.g.][and references therein]{galloway04},
that arises from overestimating the losses from neutrino emission as given by \citet{fujimoto87}.} 
\citep{goodwin19a}. 
Once the ignition depth is known, the recurrence time between the bursts can be calculated by using the equation $\Delta t_{\rm rec} =(y_{\rm ign} /\dot{m})(1 + z)$. 

It is not possible to ambiguously estimate $y_{\rm ign}$ or $\Delta t_{\rm rec}$ because of the lack of knowledge of the hydrogen fraction $\langle X\rangle$ at ignition (and hence $Q_{\rm nuc}$) in Eq. (\ref{eq:yign}). In principle it may be possible to estimate $\langle X\rangle$ based on the shape of the light curve, since that is known to be influenced by the accretion fuel \citep[e.g.][]{galloway08b}. Instead, we apply a Monte-Carlo approach and estimate $y_{\rm ign}$ and $\Delta t_{\rm rec}$ for a set of randomly generated values of $\langle X\rangle$ uniformly distributed in the range 0--0.7. Given each pair of ($\langle X\rangle$,  $\Delta t_{\rm rec}$) we can infer the accreted H-fraction $X_0$, based on the expectation that hydrogen burns via the hot-CNO cycle and will be exhausted in a time $t_{\rm CNO} =9.8(X_0/0.7)(Z_{\rm CNO}/0.02)^{-1}$ \citep[e.g.][]{lampe16}.
Since $Z_{\rm CNO}$ is also unknown, we similarly draw random values from a uniform distribution in the range 0.0--0.02.
We then apply a cut on the inferred $X_0$ values, requiring $X_0<0.75$, and compute the confidence limits on the remaining parameters. 
The inferred ignition column is $2.2_{-0.8}^{+1.4}\times10^8\,{\rm g\,cm^{-2}}$ and the average expected recurrence time is $1.8\pm0.7$~d. Such long recurrence times likely guarantee that the hydrogen in the accreted fuel is exhausted at the base prior to ignition, and explaining the fast rise and short duration of the burst. We can also infer lower (upper) limits on $X_0$ ($Z_{\rm CNO}$), although these limits are not strongly constraining; we find $X_0>0.17$ and $Z_{\rm  CNO}<0.017$ at 95\% confidence.

The inferred long recurrence time likely explains the lack of detection of other X-ray bursts during the continuous $\sim 35$~ks \xmm, $27.3$ and $29.4$~ks \nustar\ and $20$~ks \cxo\ observations, as well as
in the segmented \nicer\ observations of $93.4$~ks in total, and \swift\ observations of $17.5$~ks in total.

Taking all the results together, the type-I X-ray burst fast rise time, the $\sim 10$ s long plateau at its peak (typical for a PRE episode), the ignition depth and its related values, we conclude that the JEM-X event was triggered by unstable helium burning, after all accreted hydrogen was exhausted by  steady burning prior to the burst.

\section{Discussion}
\label{sect_disc}

When an AMXP undergoes an X–ray outburst, reaching luminosities between $10^{36-37}$ erg s$^{-1}$, most of them display very similar behaviours. The outburst lightcurves are typically characterised by a few days rise followed by a few weeks exponential- to a few days linear decay down to the quiescence level \citep[see e.g.,][]{mfb05,falanga11,falanga12,ferrigno2011,deFalcob,deFalcoa}. These AMXP outburst profiles are commonly interpreted in terms of the disk instability model, taking into account the irradiation of the disk by the central X-ray source \citep{King1998}. The outburst profile, timescale -- from a few weeks to months -- and its luminosity are well fitted within the disk instability picture \citep{Powell2007}.

The outburst profile of \psrtar\ with several reflarings can be explained as follows. During the first 60 days the companion star was able to supply the neutron star with a variable mass accretion rate, and after reaching its maximum flux level -- around 60 days since the onset of the outburst (near MJD 58379; the last reflare) -- the last episode of $\sim 25$ days of the full outburst showed behaviour similar to other AMXPs.

\begin{figure}[t]
  \centering
  \includegraphics[angle=0,width=9.0cm]{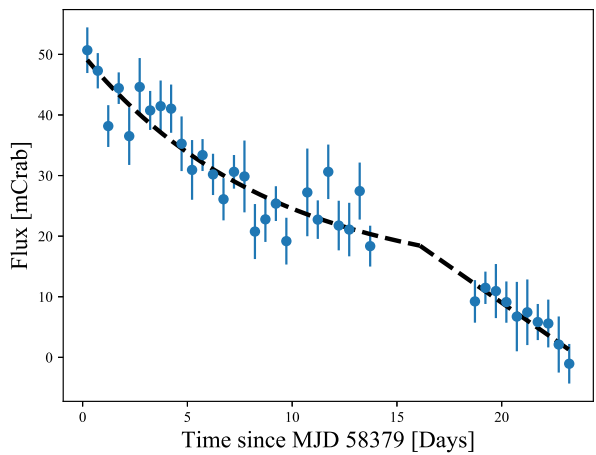}
  \caption{\psrtar\ late-time \Integ-ISGRI (20--60 keV) outburst profile (last reflare) starting from MJD 58379  with a time bin size of 0.5 d. The dashed line corresponds to the best fit in terms of an exponential profile, $F\propto e^{-t/9^{d}}$, until the break (``knee") followed by a linear decay.}
  \label{fig:outburst_decay}
\end{figure}

Its late-time outburst shape fits nicely within the framework of the disk instability model, i.e. the lightcurve decays exponentially until it reaches a break (``knee"), in our case around MJD 58395.5, after which the flux drops linearly within $\sim 7$ days to the quiescence level (see Figs. \ref{outburst_profile} and \ref{fig:outburst_decay}, a zoom-in of the last reflare). 

The knee occurs at the lowest value of the X-ray luminosity at which the outer disk edge can still be kept hot by central illuminating source. Following \citet{Powell2007} it is possible to estimate the outer disk radius in two different and independent ways. From the equation $L_{\rm X} = (L_{\rm t}-L_{\rm e})\,{\rm exp}(-(t-t_{\rm t})/\tau_{\rm e}) + L_{\rm e}$, where $L_{\rm e}$, $t_{\rm t}$ (the break time), $L_{\rm t}$, and $\tau_{\rm e}$ (exponential decay time) are all free parameters, we determined from the fit to the decay lightcurve (see Fig. \ref{fig:outburst_decay}), the outer disk radius as $R_{\rm disk}(\tau_{\rm e})= (\tau_{\rm e}\,3\nu_{\rm KR})^{1/2} \sim
3.0\times10^{10}$ cm. We adopted for the viscosity near the outer disk edge the value of $\nu_{\rm KR}= 4\times10^{14}$ cm$^2$ s$^{-1}$ \citep[see][ for more details]{King1998,Powell2007}. 

An independent estimate of $R_{\rm disk}$ can also be obtained by using the equation $R_{\rm disk}(L_{\rm t})= (\Phi\,L_{\rm t})^{1/2}$. If we use $\Phi\sim 1.3\times10^{-15}$ cm$^2$ s erg$^{-1}$ as determined from other AMXPs \citep[][]{Powell2007}, we obtain
$R_{\rm disk}(L_{\rm t})\sim (5.8-7.0)\times10^{10}$ cm, for a source distance between 7.0--8.4 kpc (see Sect. \ref{sec:burstprop}). The estimated $R_{\rm disk}(L_{\rm t})$ is larger than $R_{\rm disk}(\tau_{\rm e})$ by a factor of 2.

At variance to other AMXPs outbursts, for \psrtar\ the exponential and linear outburst decay profile occurred only after $\sim 60$ days after start of the outburst, which may produce a hotter disk, and hence a smaller $\Phi$ value. The  parameter $\Phi \propto T_{\rm h}^{-4}$ depends mainly on the irradiation temperature, $T_{\rm h}$, at the outer disc radius \citep{King1998}. An agreement of the outer disk radius, $R_{\rm disk}(L_{\rm t}) \sim 3.0\times10^{10}$ cm,  can be obtained if $\Phi$ is reduced by a factor of 4. We note, that for other systems, e.g., black hole binaries, the factor $\Phi$ is also reduced by factor of 10.  
This radius also fulfills the condition $R_{\rm circ} \lesssim R_{\rm disk}(\tau_{\rm e}) <
b_{1}$, where $R_{\rm circ} \sim 2.9 \times 10^{10}$ cm is
the circularization radius and $b_{1} \sim 1.1 \times10^{11}$
cm is the distance of the Lagrange point $L_{1}$ from the center of
the neutron star \citep[see e.g.,][]{frank02}. To estimate these values
we used a companion star mass of $\sim0.9M_{\odot}$, for a neutron star mass of 1.4 $M_{\odot}$. We note, how for this source the $R_{\rm circ}$ is close to the outer $R_{\rm disk}$ radius. 

The observed timing behaviour of \psrtar\ (see Fig. \ref{toa_residuals}) is rather complex: starting with a spin-up episode (MJD 58345--58364), followed by a frequency drop (glitch-like feature; MJD 58364--58370), a period of constant frequency (MJD 58370--58383) and concluding with irregular behaviour beyond  (MJD 58383--58408). 
This behaviour is likely related to varying accretion torques during outburst which are common in persistent disk-fed pulsars \citep{bildsten1997}. The underlying mechanisms are probably also acting in disk-fed pulsars in transient systems. The structure in the ToA residuals (see Fig. \ref{toa_residuals}) can {\it not\/} be explained by a bias model containing a bolometric flux dependent term \citep[see e.g.][]{Patruno09}, which e.g. nicely works for the ToA modelling of the 2019 outburst of SAX J1808.4--3658 \citep[see][ for flux-adjusted phase model]{bult2019}.

Our detailed pulse profile Fourier decomposition as a function of energy across the 1--50 keV band (see Fig. \ref{pp_fourier_fit}) showed that up to $\sim 9$ keV harder X-ray photons arrive earlier than soft X-ray photons.
Above this energy the lag seems to recover to smaller values.
This behaviour bears strong similarity to the picture observed for \igrone, in which Compton scattering process was proposed to play an important role in shaping the observed time-lag features \citep[see e.g.][]{Falanga07}.
Alternatively, the observed evolution of time lags can be explained by different emission pattern of the soft blackbody   and hard Comptonized photons \citep{Poutanen03}.

In this work we have analysed the broad band (0.3--300 keV) time-averaged emission spectrum of \psrtar\ adopting different spectral models among which an absorbed \compps\ model in order to facilitate comparisons with the spectra observed for other AMXPs. We obtained a coherent spectral picture with blackbody seed photon temperature $kT_{\rm bb,seed}$, electron temperature $kT_{\rm e}$ and Thomson optical depth $\tau_T$ comparable with values measured for other AMXPs. Moreover, the fitted normalisation provided an emission area radius compatible with the size of a neutron star adopting a distance of $7.6$ kpc, as derived in this work from the characteristics of the (single) type-I burst discovered in \Integ-JEM-X data. 
The size of the emission region would be significantly  smaller if the temperature of seed blackbody photons coming from below the Comptonizing slab would be larger (of about 1.5 keV) than the directly seed blackbody coming from a region outside of accretion shock, as is observed in other AMXPs \citep[e.g.][]{mfc07}. 
The total hydrogen column density $N_{\rm H}$ of $(2.09\pm0.05)\times 10^{22}$ cm$^{-2}$ derived in this work, however, is considerably less than the values derived by other groups analysing spectral data of \psrtar\ using a different combination of high-energy instruments \citep[see e.g.][]{Russell2018,Sanna2018,Nowak2019,Gusinskaia2020}. The difference can be explained by the degeneracy between the $N_{\rm H}$ and $kT_{\rm bb,seed}$ fit parameters which became clear through the inclusion of high-quality spectral data from \xmm\ RGS 1 \& 2 and EPIC-MOS2 for energies below $\sim$1 keV in this work. 
Moreover, the reconstructed total hydrogen column density in the direction of \psrtar\ is consistent with the total Galactic column density as derived from reddening maps \citep{Russell2018}, and so there is no need for an intrinsic absorption component near the source.

AMXPs are characterised by an orbital period ranging from $\sim 38$ min (\igrweak) to $\sim 11$ hrs (\igrswing), i.e. hosting hydrogen-rich main-sequence companion stars for AMXPs with $> 3$ hrs orbital periods, brown dwarfs companion stars with orbital periods between 1--2 h, or ultra-compact helium white dwarfs for orbital periods around 40 minutes \citet{Campana18}.

All the AMXPs exhibiting type-I X-ray bursts have orbital periods larger than 1 h, i.e. those with hydrogen-rich main-sequence or brown dwarf companion stars.
However, those with ultra-compact helium white dwarf companions do not exhibit type-I X-ray bursts, except \igrweak\ that showed one long helium burst, which could be explained since this source is in a faint persistent outburst since 2006.
Therefore the detection of a type-I X-ray burst from \psrtar\ is in perfect agreement with the current picture of type-I X-ray burst occurence in AMXPs.

\section{Summary}
\label{sect_summary}
In this work we have analysed all available \Integ-ISGRI (20--300 keV) data obtained from \Integ\ observations performed between Aug 10 -- Oct 11, 2018
(Revs. 1986 -- 2009) together with data from two \nustar\ (3--79 keV) ToO observations, one \xmm\ (0.3--15 keV) ToO observation and \nicer\ (0.2--12 keV) 
monitoring observations in order to obtain a complete picture of the high-energy emission properties of accretion-powered millisecond X-ray pulsar \psrtar\ 
during its 2018 outburst in both the timing- and spectral domain. 
A summary of our findings are given below:
\newcounter{sumlist}
\begin{list}%
{(\roman{sumlist})}{\usecounter{sumlist}\setlength{\rightmargin}{\leftmargin}}
\item The hard X-ray outburst profile obtained from \Integ-ISGRI 20--60 keV and \swift-BAT 20--50 keV monitoring data showed that
\psrtar\ was active for about 85 days from MJD 58320 (July 21, 2018) till MJD 58405 (Oct. 14, 2018). The outburst profile showed multiple re-flarings, and 
during the last re-flaring we detected in \Integ\ JEM-X data a bright thermonuclear burst from \psrtar\ on MJD 58380.96358 (MJD TT; Sept. 19, 2018 23:07:33).

\item Under the assumption of Eddington luminosity limited PRE burst emission, and assuming an anisotropy factor of $\xi_{\rm b} = 0.71$, we estimated the source distance as $d=7.6\pm0.7$ kpc for a He burst.

\item From \nicer\ monitoring observations we updated the orbital parameters of the binary system and subsequently performed pulse a time-of-arrival analysis
in order to obtain phase-coherent timing models.

\begin{enumerate}
   \item We could derive the timing parameters ($\nu,\dot\nu$) accurately for two time segments across MJD 58345--58405.
        For segment-1 (MJD 58345--58364) we measured a spin-up rate of $\dot\nu=(4.07\pm0.79)\times 10^{-14}$ Hz s$^{-1}$, indicating a 
        net gain of angular momentum from the accretion flow. During the other segment (\#3; MJD 58370-58383) the measurements were
        consistent with the assumption of a constant spin frequency. Between and beyond these segments the timing behaviour appeared 
        to be noisy/erratic.
   
   \item Application of these timing models in pulse-phase folding procedures using \nicer, \xmm\ EPIC-Pn, (corrected) \nustar\ and \Integ-ISGRI data
         yielding pulse-phase distributions across the 0.3--300 keV band showed that pulsed emission had been detected significantly between 1 and 120 keV.
         
   \item The decomposition of the \nicer, \xmm\ EPIC-Pn and \nustar\ pulse-phase distributions across the 1--50 keV band in terms of 3 harmonics
         showed that the (background corrected) pulsed fraction of the fundamental component increases from $\sim$10\% to $\sim$17\% when going from
         $\sim$1.5 keV to $\sim$4 keV, beyond where it more or less saturates. The pulsed fraction for the other components is relatively constant:
         $\sim$6\% and $\sim$1--2\% for the first and second overtone, respectively.
         
         The phase-angle as function of energy showed a decreasing trend from $\sim$1.5 keV till $\sim$10 keV for both the fundamental and first overtone. This explains
         that the harder pulsed photons (at 10 keV) arrive $\sim$95\,$\mu$s earlier than softer pulsed photons (at 1.5 keV).
   
\end{enumerate}

\item The total broad-band emission spectrum of \psrtar\ using \xmm\ EPIC-Pn, MOS-2, RGS 1\&2, \nustar\ and \Integ-ISGRI data (0.3--150 keV) was fitted adequately with
a thermal Comptonization model (\compps) in slab geometry absorbed through an intervening column of $N_{\rm H}=(2.09\pm0.05)\times 10^{22}\, $\,cm$^{-2}$. 
Other best-fit parameters for this model were: a blackbody seed photon temperature $kT_{\rm bb, seed}$ of $0.64\pm 0.02$ keV, electron temperature $kT_{\rm e}=38.8\pm 1.2$ keV and Thomson optical depth $\tau_{\rm T}=1.59\pm 0.04$, while the normalization translated to an emitting radius of $11.3 \pm 0.5$ km assuming a distance of 7.6 kpc.

\end{list}

\begin{acknowledgements}
This research has made use of data obtained from the High Energy Astrophysics 
Science Archive Research Center (HEASARC), provided by NASA's Goddard Space Flight Center.
We have extensively used NASA's Astrophysics Data System (ADS). 
SST, IAM and JP were supported by the grant 14.W03.31.0021 of the Ministry of  Science and Higher Education  of  the  Russian Federation.
\end{acknowledgements}

\bibliographystyle{aa}
\bibliography{pulsars}

\end{document}